\documentclass[hyper,a4paper,12pt,oneside]{JHEP3}
\usepackage{amsmath,amscd}
\usepackage{amsfonts}
\usepackage{amssymb}
\usepackage{graphicx,shortvrb}
\usepackage{epsfig}
\usepackage{newlfont}
\usepackage[vcentermath]{youngtab}

\makeatletter
\renewcommand{\subsubsection}{\@startsection{subsubsection}{3}{0mm}{-\baselineskip}{0.5\baselineskip}{\normalfont\normalsize\it}}
\makeatother

\newcommand{\csp}[1]{\hspace{#1em},\hspace{#1em}}

\newcommand{\rar}[1]{\hspace{#1em}\Rightarrow\hspace{#1em}}
\newcommand{\N}{\mathcal{N}}
\newcommand{\Op}{\mathcal{O}}
\newcommand{\Qp}{\mathcal{Q}}
\newcommand{\Bp}{\mathcal{B}}

\newcommand{\Jdp}{\mathcal{J}{\kern-.27em d}}
\newcommand{\Adp}{\mathcal{A}{\kern-.18em d}}
\newcommand{\Jdet}{\mathrm{Jdet}}
\newcommand{\ads}[1]{AdS$_#1$}

\newcommand{\bps}[1]{$1/{#1}$-BPS}

\newcommand{\pdf}[2]{\frac{\partial#1}{\partial#2}}
\renewcommand{\l}{\lambda}
\newcommand{\eps}{\epsilon}
\DeclareMathOperator{\SU}{SU} \DeclareMathOperator{\SO}{SO}
\DeclareMathOperator{\tr}{Tr}
\newcommand{\pl}[2]{Pl\left[#1,\,#2\right]}

\newcommand{\abs}[1]{\left\vert#1\right\vert}

\newcommand{\ket}[1]{\left\vert #1\right\rangle}

\newcommand{\zbra}[1]{\Bigl\langle\relax{\kern-.4em}\Bigl\langle#1\Bigr\vert}
\newcommand{\zinnp}[2]{\Bigl\langle\relax{\kern-.4em}\Bigl\langle #1\Bigm\vert #2\Bigr\rangle}            
\newcommand{\zbok}[3]{\Bigl\langle\relax{\kern-.4em}\Bigl\langle #1\Bigl\vert #2\Bigr\vert#3\Bigr\rangle} 

\preprint{\small UPR-1149-T\\\small WIS/04/06-APR-DPP}

\title{A Fermi Surface Model for Large Supersymmetric \ads5 Black Holes}
\author{Micha Berkooz$^1$, Dori Reichmann$^1$, and Joan Sim\'on$^2$\\
$^{1}$Department of Particle Physics, The Weizmann Institute of
Science, Rehovot 76100, Israel\\
$^{2}$David Rittenhouse Laboratories, University of Pennsylvania,
Philadelphia, PA 19104, U.S.A.}
\date{\today}
\abstract{We identify a large family of 1/16 BPS operators in $\N=4$
SYM that qualitatively reproduce the relations between charge,
angular momentum and entropy in regular supersymmetric \ads5 black
holes when the main contribution to their masses is given by their
angular momentum.
}
\begin{document}
\setcounter{tocdepth}{1}

\section{Introduction}\label{z-tit-int}

One of the successes of string theory is the statistical mechanics
derivation of the entropy of supersymmetric black holes
\cite{Strominger:1996sh}. In particular, within the context of the
AdS/CFT correspondence
\cite{Maldacena:1997re,Witten:1998qj,Gubser:1998bc,Aharony:1999ti},
one can carry out detailed computations of black hole properties and
compare them to results from the dual conformal field theory.

One of the most heavily studied cases of the AdS/CFT is the duality
between type IIB string on $AdS_5\times S^5$ to $\N=4$ SYM. In this
case, however, black hole entropy counting has turned out to be
tricky. For large non-supersymmetric black holes $S= c N^2 T^3$
\cite{Witten:1998zw}. The power of $T$ is determined by dimensional
analysis, $N^2$ is the free theory scaling of degrees of
freedom\footnote{in 3+1 dimensions, the N scaling survives to strong
't Hooft coupling. This does not necessarily happen in other
dimensions.}, but the coefficient, $c$, can not be computed reliably
\cite{Gubser:1996de,Balasubramanian:2005kk}. In the arena of
supersymmetric R-charged configurations, there are no honest black
holes in the $1/2$ to \bps8 sector states, i.e, there is less than
$N^2$ worth of entropy, and so their horizons are Planck scale.

In the current paper we take some first steps towards understanding
the AdS/CFT correspondence in cases with $1/16$ SUSY. With this
amount of supersymmetry, there is a rich spectrum of genuine black
holes with smooth horizons and non vanishing angular momentum
\cite{Gutowski:2004ez,Gutowski:2004yv,Chong:2005da,Kunduri:2006ek}.
Our goal is to, first, find their field theory duals (forcing us to
understand the addition of angular momentum), and second, to count
them. In this paper we explain some structures which are key to the
construction of the $1/16$ operators in the CFT. Our construction is
based on the filling of fermi surfaces, and it is similar in spirit
to the construction in terms of free fermions for the $1/2$ - $1/8$
BPS states
\cite{Corley:2001zk,Berenstein:2004kk,Lin:2004nb,Berenstein:2005aa}.
Using this structure we reproduce the scalings between angular
momentum, charge and entropy of $1/16$ SUSY black holes, up to
coefficients of order 1. Furthermore, since the fermi surface is
multi-dimensional, it posses a complicated morphology. This suggests
that additional types of black holes might be constructed, with an
equally complicated bulk morphology.

Asymptotically supersymmetric \ads5 black holes were originally
constructed in \cite{Gutowski:2004ez,Gutowski:2004yv} and later
generalized in \cite{Chong:2005da,Kunduri:2006ek}. These black holes
carry both angular momenta\footnote{In the supergravity literature,
it is customary to use ~$J_{1,2}=J\pm \bar J$, where $J_{1,2}$ are
angular momenta on two orthogonal 2-planes.} $\{J,\,\bar J\}$ under
$\SU(2)_L\times\SU(2)_R$ and $\SO(6)$ R-charges
$\{Q_1,\,Q_2,\,Q_3\}$. Their mass is given by the BPS equation:
\begin{equation}
  M = \frac{2J}{l} +  Q_1 + Q_2 + Q_3\,.
 \label{eq:bound}
\end{equation}
The $Q_i$'s are taken to be of dimension 1, and $l$ is the \ads5
radius.

There are two natural scaling regimes to consider according to
whether the R-charge or the angular momenta is large. In this note,
we study the regime in which the black hole mass is dominated by the
angular momenta. For simplicity, we focus on black holes having
three equal $\SO(6)$ R-charges $Q_1=Q_2=Q_3=\frac{Q}{2\sqrt3}$.
Black holes in this regime exhibit different angular momentum-charge
relations depending on their right handed angular momentum, $\bar J$
(which does not appear in the BPS formula). The two scaling
behaviors that we will be interested in
\begin{align}
    J/N^2 \sim& (Q/N^2)^{3/2},&\text{if}\qquad \bar J\simeq&0\csp1 Q>>N^2\,,&\cr
    J/N^2 \sim& (Q/N^2)^2,&\text{if}\qquad \bar J\simeq&J\csp1 Q>>N^2\,.&
     \label{sclngabs}
\end{align}
We identify the correct short representations of the superconformal
group, and construct highest weight chiral operators in these
representations whose quantum numbers not only satisfy the BPS bound
\eqref{eq:bound}, but satisfy the scaling relations
\eqref{sclngabs}.

Our models rely on shells of fermions, forming a fermi sea. It is
easy to motivate the need for such a fermi sea when describing
operators satisfying \eqref{sclngabs}. Consider bringing together
two such black holes in AdS. For simplicity we focus on the case
$J=\bar J$. Each black hole has charge $Q_i$ , $i=1,2$ , and angular
momentum $J_i=\bar J_i\propto\frac{Q_i^2}{N^2}$. Suppose that we
place the black holes with no relative angular momentum. In this
case they cannot merge to form a new black hole with $Q=Q_1+ Q_2$
since there is not enough angular momentum for the latter. The black
holes have to remain distinct from each other, suggesting a sort of
fermi exclusion principle. To have the black holes fuse we need to
provide more angular momentum to the system.

In the field theory the interpretation is the following. Let us
consider the OPE of two \bps{16} operators that correspond to one of
the microstates of these black holes. Focusing on the \bps{16}
operator in the expansion with total charge $Q_1+Q_2$, and denoting
\begin{equation}
    \delta J=J(Q_1+Q_2)-J(Q_1)-J(Q_2)\propto \frac{Q_1Q_2}{N^2}
\end{equation}
then the $Q_1+Q_2$ \bps{16} operator appears in the OPE as a regular
term with a power $x^{2\delta J}$ in front. The $Q_1$ and $Q_2$
operators therefore cannot be at the same point in space-time. This
is reminiscent of two fermions OPE, $\psi (0) \psi(x) \sim x
(\psi(0)\partial\psi(0))$ and its N-species generalization
\begin{equation}
    \left(\prod_{i=1}^N \psi_i(0)\right)\left(\prod_{j=1}^N \psi_j(x)\right)\sim x^N
    \left(\prod_{i=1}^N
    \psi_i(0)\partial\psi_i(0)\right)
\end{equation}
The rest of the paper explains what are the relevant fermions and
their precise structure.

It is important for us to work in the interacting theory. Indeed in
\cite{Kinney:2005ej}, the spectrum of \bps{16} in the free theory
was computed, and was found not to satisfy relations of the type
\eqref{sclngabs}. However if one imposes this relation, although
there are too many operators the entropy is larger only by a
numerical coefficient. We will work in the interacting theory and
establish the origin of \eqref{sclngabs}, but again up to a
numerical coefficient\footnote{In \cite{Kinney:2005ej}, the entropy
of small, charge dominated black holes, was counted using D-branes.
But this counting did not explain the $J(Q)$ relations that we
will.}.

The fact that we did not obtain the correct numerical coefficient in
$J(Q)$ is not surprising since, as will be seen, we have focused
only on a subset of possible fundamental fields and fermi surface
configurations. Clearly, it will be important to generalize the
operators in both avenues in order to enumerate all the
possibilities.

The paper is organized as follows. In section 2 we discuss the two
scaling of large black holes that we will be interested in and which
give \eqref{sclngabs}. In section 3 we set up some field theory
aspects that are needed for our model. In section 4 we discuss the
heuristic model of fermi surfaces and reproduce qualitative aspects
of $J(Q)$ and entropy, we also construct a class of BPS operators.
Section 5 contains elaborations of the basic construction of section
4. Section 6 contains some conclusions and directions for future
research.

\section{Large Supersymmetric \ads5 Black Holes}\label{z-tit-large}

Explicit constructions of supersymmetric black holes in global
\ads5$\times S^5$ with regular finite horizons were found in
\cite{Gutowski:2004ez,Gutowski:2004yv,Chong:2005da,Kunduri:2006ek}.
These spacetime configurations were obtained either by solving the
corresponding gauged supergravity equations of motion and
supersymmetry constraints, or by studying the BPS limits of
non-extremal rotating R-charged \ads5 black holes.

As a result of this analysis, one learns that these black holes can
carry all possible charges $\{J,\,\bar J,\,Q_1,\,Q_2,\,Q_3\}$
appearing in the maximal compact subgroup of $\SO(2,4)\times\SO(6)$,
and that they preserve $1/16$ of the total supersymmetry. $J$ and
$\bar J$ stands for the angular momentum on the $S^3$ in \ads5. The
set $\{Q_i,\,i=1,2,3\}$ stands for the angular momenta on the
transverse $S^5$ and spans the $\SO(2)\times\SO(2)\times\SO(2)$
Cartan subalgebra of $\SO(6)$\footnote{These $\SO(6)$ charges are
the ones appearing naturally in supergravity. The relation between
these and the $\SU(4)$ R-charges in the dual $\N=4$ SYM is discussed
in the appendix \ref{z-tit-apd-a}.}.

In this work, we focus on supersymmetric \ads5 black holes with
equal R-charges\footnote{Taking the three R-charges equal
$Q_1=Q_2=Q_3=\frac{Q}{2\sqrt3}$ in the notations of
\cite{Gutowski:2004yv}.} $\Qp=Ql/\sqrt{3}$, and two independent
angular momenta \cite{Chong:2005da} :
\begin{subequations}
\begin{align}
    J+\bar J=& J_1=N^2\frac{(a+b)(2a+b+ab)}{4(1-a)^2(1-b)}\,,\\
    J-\bar J=& J_2=N^2\frac{(a+b)(a+2b+ab)}{4(1-a)(1-b)^2}\,,\\
    \Qp=&Q l/\sqrt3=N^2\frac{a+b}{(1-a)(1-b)}\,,\\
    Ml=&2|J|+\frac{3}{2}\Qp\,.
  \label{BPS-cond}
\end{align}
\end{subequations}
The last equation is a manifestation of the supersymmetry of the
system since it corresponds to a standard BPS equation relating the
mass of the state with its charges. More precisely the exact formula
is given in \eqref{BPS-formula} and it differs from \eqref{BPS-cond}
by a factor which is invisible in the supergravity approximation.

Since all these black holes have a finite horizon area, we can
associate a non-vanishing entropy to them through the
Bekenstein-Hawking relation :
\begin{equation}
  S_{BH}=\pi N^2\frac{(a+b)\sqrt{a+b+ab}}{(1-a)(1-b)}= \pi\,\Qp\,\sqrt{a+b+ab}\,.
 \label{eq:entropy}
\end{equation}
Thus, the gravitational description of these black holes is
characterized by three independent parameters $\{N,\,a,\,b\}$. As
usual, $N$ fixes the flux of the RR five form.

There are three different scaling limits to consider depending on
whether the main contribution to the mass is given by the angular
momentum sector ($|J|\gg \Qp$), the R-charge sector ($\Qp\gg|J|$) or
both ($|J|\sim \Qp$). In the following, we shall concentrate in the
limit
\begin{equation}
    \frac{|J|}{N^2}\gg\frac{\Qp}{N^2} \gg 1\,, \qquad
    Ml\approx 2|J|\,.
  \label{eq:uslimit}
\end{equation}
As already emphasized in the introduction, this is a limit in which
we expect to learn something fundamentally new about the physics of
the system since it focuses on the angular momentum sector of the
black hole. If we were to consider the first limit, it would be
natural to adopt a description in terms of fluctuations on top of
giant gravitons, as in \cite{Kinney:2005ej}.

There are two inequivalent ways of achieving the limit
\eqref{eq:uslimit}. These are obtained by scaling either both
parameters $\{a,\,b\}$ to their extremal values $(a,\,b\to1)$\footnote{To avoid the existence of the so called theta horizons and closed timelike curves, the parameters $\{a,\,b\}$ satisfy the constraint $|a|,\,|b| < 1$.}, or
just one of them :
\begin{itemize}
    \item Scaling I : $J \to\infty,\ {\bar J}/J\to\gamma<1$
    This corresponds to studying the scaling
\begin{equation*}
      a = 1-\alpha\epsilon\,, \qquad b=1-\beta\epsilon\,, \qquad \epsilon\to 0^+\,
      \qquad(\alpha,\beta>0).
\end{equation*}
The angular momentum of the black hole\footnote{The reverse
    case $J<{\bar J}$ is just the parity transformation of this
    case. As derived in \cite{Dobrev:1985qv} unitarity implies that if
    the operator is annihilated with a combination of the $Q$'s then
    ${\bar J}\leq J$.} and entropy are given by
    \begin{subequations}\begin{align}
        & J \to \frac{N^2\sqrt{2}}{\sqrt{1-\gamma^2}}\left(\frac{\Qp}{
        N^2}\right)^{3/2}\label{JtoQ_ratio}\\
        &S_{BH} \to \sqrt3\,\pi\,\Qp\,,\label{entropy-sugra}
    \end{align}
    \end{subequations}
where $\gamma = (1-\alpha/\beta)/(1+\alpha/\beta)$ and it is smaller
than one by construction. We mainly focus on the $\gamma=0$
$(\alpha=\beta)$ case where the solution is $\SU(2)_R$ invariant.
    \item Scaling II :  $J\to\infty,\ {\bar J}/J\to 1$.
This corresponds to scaling only one parameter
\begin{equation*}
      a = 1-\epsilon\,, \qquad b~~\text{fixed}~ (< 1~)\,, \qquad \epsilon\to 0^+\,.
\end{equation*}
The angular momentum and entropy  of the system behave as
 \begin{subequations}\begin{align}
        \label{JbarJtoQ ratio}
        &J=\bar J\to\frac{N^2}{4}\,(1-b)\left(\frac{\Qp}{
        N^2}\right)^{2}\\
        &S_{BH}\to\pi\,\sqrt{1+2b}\,\Qp\,.
        \label{entropy-sugra-jbar j}
    \end{align}\end{subequations}
It is important to keep in mind that even though $\bar J\to J$,
their difference is non-vanishing
\begin{equation}\label{JminBarJ-sugra}
  J-\bar J \to \frac{1}{4}\,\frac{1+3b}{1-b}\,\Qp
  \quad\left(\ll J\,,\,\bar J\right)\,.
\end{equation}
\end{itemize}

Relations \eqref{JtoQ_ratio} and \eqref{JbarJtoQ ratio} are
particular examples of the general statement non-linear constraints
among the global charges of the black hole. They come from the
resolution of the supergravity equations of motion and supersymmetry
constraints. They are not implied by the superconformal symmetry of
the theory, as we shall review below. One of our goals is to provide
an explanation for these scaling relations in the dual $\N=4$ SYM.

\section{Field Theory Aspects}\label{z-tit-field}

In this section, we identify the superconformal representations
associated with these black holes. We also describe the main
building blocks of the chiral operators we construct later on.

Let us first introduce some notation. We are using the conventions
for $\N=4$ SYM from \cite{Ryzhov:2001bp}. The component fields of
the $\N=4$ super-multiplet are denoted by :
\begin{itemize}
  \item[(i)] $F_{\alpha\beta}$ and $\bar F_{\dot\alpha\dot\beta}$ for the gauge fields
  \item[(ii)] $\l_{\alpha i}$ and $\bar \l^{i}_{\dot\alpha}$ for the gauginos
  \item[(iii)] $M_{ij}$ for the scalars
\end{itemize}
Undotted ($\alpha$), dotted ($\dot\alpha$) greek indices and latin
($i$) indices stand for $\SU(2)_L\times\SU(2)_R\times\SU(4)$
symmetry indices, respectively. Left-handed fermions transform in
the anti-fundamental representation of the $\SU(4)$ R-symmetry group
whereas right-handed fermions transforms as a fundamental. Scalars
transform in the anti-symmetric 2-tensor representation of $\SU(4)$
and obey the reality condition:
\begin{equation*}
    (M_{ij})^\dag=\bar M^{ij}=\frac12\eps^{ijkl}M_{kl}.
\end{equation*}
The $\N=4$ supersymmetry transformations are:
\begin{align}
%
    \delta\,M_{jk} &
    =\zeta_j\l_k-\zeta_k\l_j+\eps_{jklm}\bar\zeta^l\bar\l^m\,,\cr
    \delta\,\l_j &
    =F\cdot\zeta_j+2iD M_{jk}\cdot\bar\zeta^k-2i[M_{jk},\bar M^{kl}]\zeta_l\,,\cr
    \delta\,F &
    =-i\zeta_j\cdot D\bar\l^j+iD\l_j\cdot\bar\zeta^j\,,
\end{align}
where $D$ is the gauge covariant derivative and we ignored all
$\SU(2)_L\times\SU(2)_R$ and $\SU(N)$ indices to simplify the
presentation. The notation we are using has $g_{YM}$ hidden in the
definition of the gauge potential. Thus, the free-field limit $(g_{YM}=0)$
is equivalent to removing all commutation relations.

Our notations seem to "jump" at $g_{YM}^2=0$. However, the counting
of states with given R-charge and angular momentum was carried out
in \cite{Kinney:2005ej} for $g_{YM}^2=0$ with the result that the
free theory has too many of them. We expect that the number of
operators will change between $g_{YM}^2=0$ and $g_{YM}^2\not=0$ and
hence it is natural to work in notations adapted to the
latter\footnote{In fact, it is an interesting problem whether the
spectrum of 1/16 operators changes for other value of $g_{YM}^2$.
The results that we present in the rest of the paper suggests that
they do not.}.

A detailed analysis of short and semi-short $\SU(2,2|4)$
superconformal representations is presented in \cite{Dolan:2002zh}.
Here, we follow their notations. Highest weights representations of
this type are classified by six quantum numbers. One of them,  the
conformal dimension $\Delta$, is always determined by the shortening
of the representation. The information regarding the other five is
given by
\begin{equation*}
  [k,\,p,\,q]_{J,\,\bar J}
\end{equation*}
where $[k,\,p,\,q]$ stands for the Dynkin labels of the $\SU(4)$
R-charges\footnote{A representation of $\SU(4)$ with highest weight
state having  Dynkin labels $[k,\,p,\,q]$ can be represented by a
Young-Tableau with $k$ columns of height 3, $p$ columns of height 2
and $q$ columns of height 1.}. The relations between the highest
weights vector in these representations and the charges given before
(as reviewed in appendix \ref{z-tit-apd-a}) are\footnote{We are
using the similar notation to describe the Dynkin labels of the
representation and the three abelian R-charges. We use the square
brackets $[k,\,p,\,q]$ whenever we refer to the representation,
while we use round brackets $(k,p,q)$ for the weights.}:
\begin{equation}
    Q_1=\frac{k+2p+q}{2 l}\,,\qquad
    Q_2=\frac{k+q}{2 l}\,,\qquad
    Q_3=\frac{k-q}{2 l}\,.
 \label{eq:rchargedict}
\end{equation}

There are two families of \bps{16} states $\{c^{1/4},\,\bar
c^{1/4}\}$ which are conjugate to each other \cite{Dolan:2002zh}.
Highest weight states $|k,\,p,\,q;\,J,\,\bar J>$ belonging to the
$c^{1/4}$ representation satisfy the BPS condition:
\begin{equation}\label{supercharge}
  |k,\,p,\,q;\,J,\,\bar J>\,\in\,c^{1/4}\,\, \Leftrightarrow \,\, \left(Q^1_2 - \frac{1}{2J+1}\,J_-\,Q^1_1\right)\,|k,\,p,\,q;\,J,\,\bar J>  = 0\,,
\end{equation}
where $J_-$ stands for the lowering operator in $\SU(2)_L$ and
$Q^i_\alpha$ are supercharge generators. The supercharges
$Q^1_\alpha$ transform in the representation:
\begin{equation*}
  Q^1_\alpha \sim [1,\,0,\,0]_{\pm 1/2,0}\,.
\end{equation*}
An equivalent characterization of these representations can be given
in terms of null states:
\begin{gather}
    c^{1/4}:\quad[k,p,q]_{(J,\bar J)}\xrightarrow{~Q~}[k+1,p,q]_{(J-\frac12,\bar
    J)}\quad\text{null}\cr
    \bar c^{1/4}:\quad[k,p,q]_{(J,\bar J)}\xrightarrow{~\bar Q~}[k,p,q+1]_{(J,\bar
    J-\frac12)}\quad\text{null},
\end{gather}
For the $[k,0,0]$ representation, which we are interested in, the
expression is:
\begin{equation}\label{supercharge-covarint}
    \eps^{\alpha\beta_{1}}\left[Q^{(i_{0}}_{\alpha},\Op^{i_1,\ldots i_k)}_{(\beta_1,\ldots\beta_{2J})(\dot\beta_1,\ldots\dot\beta_{2\bar
    J})}\right\}=0.
\end{equation}
here $\Op$ stands for the primary operator in the $c^{1/4}$
multiplet\footnote{We are freely using the state-operator mapping to
transfer between operators in $\N=4$ SYM on $R^{1,3}$ and states of
$\N=4$ SYM on $R^{1}\times S^3$}, i.e $\Op$ is annihilated by all
superconformal supercharges. The symmetrization of the indices
$(i_0,\,i_1,\,\dots i_k)$ in \eqref{supercharge-covarint} ensures
that we pick the highest weight state with R-charge k+1, whereas the
anti-symmetrization in $\epsilon^{\alpha\beta_1}$ picks the state
with $\SU(2)_L$ angular momentum $J-1/2$. Finally, the conformal
dimension $\Delta$ of the primary operators in the multiplet is
given by the BPS formula\footnote{Remember, that all primary
operators, with the same charges, satisfy the bound
$\Delta\geq\Delta[BPS]$. The bound is saturated for BPS primary
operators. Non-vanishing operators with lower dimensions are
manifestly descendants}:
\begin{gather}\label{BPS-formula}
    \Delta[c^{1/4}]=2+2J+\frac32k+p+\frac12q\,,\cr
    \Delta[\bar c^{1/4}]=2+2\bar J+\frac12k+p+\frac32q\,.
\end{gather}
Notice that the above differ by $2$ with the conformal dimension
$(Ml)$ derived from supergravity. As mentioned before, this constant
factor is unobservable in the gravity regime where all charges are
generically taken to be large to ensure a reliable classical
spacetime description.

The Young tableau corresponding to the $Q_1=Q_2=Q_3$ operators
($p=q=0$) is:
\begin{equation*}
    \underbrace{\Yboxdim12pt\yng(4,4,4)\quad.\quad.\quad.\quad\Yboxdim12pt\yng(1,1,1)}_{k}
\end{equation*}
It should now be apparent that the non-linear constraints
\eqref{JtoQ_ratio} and \eqref{JbarJtoQ ratio} derived in
supergravity are rather non-trivial. As far as the superconformal
algebra is concerned all values of $J$, $\bar J$ and $k$ are
allowed. Our goal is to explain the details of $J(Q)$ dependence
based on the details of $\N=4$ SYM.

The global charges carried by the fundamental degrees of freedom in
the $\N=4$ super-multiplet, in the conventions introduced above, are
summarized in table-\ref{tab:fcharge}. $\Delta_{exc}$ stands for the
excess dimension compared to the global part of the BPS formula
(without the offset '2'):
\begin{equation}
    \Delta_{exc}=\Delta-2J-\frac32k-p-\frac12q
\end{equation} \TABLE{
\begin{tabular}{|c|c|c|c|c|} \hline
              & ~~~$(J,\,\bar J)$~~~& $~~~(k,\,p,\,q)~~~$ &~~~$\Delta$~~~&$~~\Delta_{exc}$~~\\
\hline
 $M_{12},\,M_{13},\,M_{14}$  &  $(0\,,\,0)$   &   $(0,\,-1,\,0),\,(-1\,1,\,-1),\,(-1,\,0,\,1)$  & $1$  & $2$    \\
 $\bar M^{12},\,\bar M^{13},\,\bar M^{14}$  &  $(0\,,\,0)$   &   $(0,\,1,\,0),\,(1,\,-1,\,1),\,(1,\,0,\,-1)$  & $1$  & $0$    \\
 $\bar\lambda^1$ & $(0,\,1/2)$ & $(1,\,0,\,0)$ & $3/2$ & $0$ \\
 $\bar\lambda^2,\,\bar\l^3,\,\bar\l^4$ & $(0,\,1/2)$ & $(-1,\,1,\,0),\,(0,\,-1,\,1),\,(0,\,0,\,-1)$ & $3/2$ & $2$ \\
 $\lambda_1$ & $(1/2,\,0)$ & $(-1,\,0,\,0)$ & $3/2$ & $2$ \\
 $\lambda_2,\,\l_3,\,\l_4$ & $(1/2,\,0)$ & $(1,\,-1,\,0),\,(0,\,1,\,-1),\,(0,\,0,\,1)$ & $3/2$ & $0$ \\
 $F_{\alpha\beta}$ & (1,\,0) & $(0,\,0,\,0)$ & $2$ & $0$ \\
 $\bar F_{\dot\alpha\dot\beta}$ & $(0,\,1)$ & $(0,\,0,\,0)$ & $2$ & $2$ \\
 $D_{\alpha\dot\alpha}$ & $(1/2,\,1/2)$ & $(0,\,0,\,0)$ & $1$ & $0$ \\
\hline
\end{tabular}
\label{tab:fcharge}\caption{Fundamental Fields.
$D_{\alpha\dot\alpha}$ is the covariant derivative}}

We will be interested in the following  building blocks (all the
$\SU(2)_L\times\SU(2)_R$ indices are symmetrized):
\begin{subequations}
\begin{align}
    A^{(I)\,i}_{(\beta_1,\ldots\beta_{I})(\dot\beta_1,\ldots\dot\beta_{I+1})}&\equiv D_{\beta_1\dot\beta_1}\cdots
    D_{\beta_I\dot\beta_I}\bar\l^i_{\dot\beta_{I+1}}
    \label{A-op-def}\\
    B^{(I)\,ij}_{(\beta_1,\ldots\beta_{I})(\dot\beta_1,\ldots\dot\beta_{I+1})}&\equiv D_{\beta_1\dot\beta_1}\cdots
    D_{\beta_I\dot\beta_I}\bar M^{ij}
    \label{B-op-def}\\
    C^{(I)}_{(\beta_1,\ldots\beta_{I})(\dot\beta_1,\ldots\dot\beta_{I+2})}&\equiv D_{\beta_1\dot\beta_1}\cdots
    D_{\beta_I\dot\beta_I}\bar F_{\dot\beta_{I+1}\dot\beta_{I+2}},
    \label{C-op-def}\\
    E^{(I+1)}_{(\beta_1,\ldots\beta_{I+1})(\dot\beta_1,\ldots\dot\beta_{I})i}&\equiv D_{\beta_1\dot\beta_1}\cdots
    D_{\beta_{I}\dot\beta_{I}}\l_{\beta_{I+1} i}
    \label{E-op-def}\\
    G^{(I+2)}_{(\beta_1,\ldots\beta_{I+2})(\dot\beta_1,\ldots\dot\beta_{I})i}&\equiv D_{\beta_1\dot\beta_1}\cdots
    D_{\beta_{I}\dot\beta_{I}}F_{\beta_{I+1}\beta_{I+2}}
    \label{G-op-def}
\end{align}
\end{subequations}
The global charges of these building blocks are summarized in
table-\ref{tab:dcharge}, where the $[k,p,q]$ are the Dynkin labels
of the representations and the excess dimension is calculated for
the highest weight. The transformation properties of these operators
under the action of left-handed supercharges $Q^i$ are as follows:
\begin{subequations}
\begin{align}
    \left\{Q^i,A^{(I)j}\right\}=&-2\sum_{m=1}^{I}\binom{I}{m}\eps\left\{A^{(m-1)i},A^{(I-m)j}\right\}+\cr
    &-2i\sum_{m=1}^I\binom{I}{m}\eps\left[C^{(m-1)},B^{(I-m)ij}\right]-2iB^{(I+1)ij}
    \label{A-tran-ex}\\
    \left[Q^i,B^{(I)jk}\right]=&
    -2\sum_{m=1}^{I}\binom{I}{m}\eps\left[A^{(m-1)i},B^{(I-m)jk}\right]+\cr
    &+\varepsilon^{ijkl}\left(\frac12\varepsilon_{i'j'k'l'}\sum_{m=1}^I\binom{I}{m}\eps\left[A^{(m-1)k'},B^{(I-m)i'j'}\right]
    +E^{(I+1)}_{l}\right)
    \label{B-tran-ex}\\
    \left[Q^i,C^{(I)}\right]=&-2\sum_{m=1}^{I}\binom{I}{m}\eps\left[A^{(m-1)i},C^{(I-m)}\right]
    -2iA^{(I+1)i}
    \label{C-tran-ex}
\end{align}
\begin{align}
    \left\{Q^i,E^{(I+1)}_j\right\}=&-2\sum_{m=1}^{I+1}\binom{I+1}{m}\left\{A^{(m-1)i},E^{(I+1-m)}_j\right\}+\cr
    &-i\varepsilon_{jkln}\sum_{m=1}^{I+1}\binom{I+1}{m}\left[B^{(m-1)ln},B^{(I+1-m)ki}\right]
    +\delta^i_jG^{(I+2)}
    \label{E-tran-ex}\\
    \left[Q^i,G^{(I+2)}\right]=&-2\sum_{m=1}^{I+2}\binom{I+2}{m}\left[A^{(m-1)i},G^{(I+2-m)}\right]+\cr
    &-2i\sum_{m=1}^{I+2}\binom{I+2}{m}\left[E^{(m-1)}_j,B^{(I+2-m)ji}\right]
    \label{G-tran-ex}
\end{align}
\end{subequations}
\TABLE{
\begin{tabular}{|c|c|c|c|c|}
\hline
              & ~~~$(J,\,\bar J)$~~~& $~~~[k,\,p,\,q]~~~$ &~~~$\Delta$~~~&$~~\Delta_{exc}$~~\\
\hline
 $A^{(I)\,,i}$  &  $(I/2\,,\,I/2+1/2)$   &   $[1,\,0,\,0]$  & $I+3/2$  & $0$    \\
 $B^{(I)\,ij}$  &  $(I/2\,,\,I/2)$       &   $[0,\,1,\,0]$  & $I+1$    & $0$      \\
 $C^{(I)}$   &  $(I/2\,,\,I/2+1)$     &   $[0,0,0]$  & $I+2$    & $2$      \\
 $E^{(I+1)}_i$ &  $(I/2+1/2\,,\,I/2)$   &   $[0,\,0,\,1]$  & $I+3/2$  & $2$      \\
 $G^{(I+2)}$ &  $(I/2+1\,,\,I/2)$     &   $[0,0,0]$  & $I+2$    & $0$      \\
\hline
\end{tabular}
\label{tab:dcharge}\caption{Building Blocks}} In the above
expressions, $\SU(2)_L\times\SU(2)_R$ indices are hidden ($\eps$ are
$\eps_{\alpha\beta}$'s), for example \eqref{C-tran-ex} reads,
\begin{multline}
    \left[Q^i_{\alpha_{I+1}},C^{(I)}_{(\alpha_1,\ldots\alpha_I)(\dot\alpha_1,\ldots\dot\alpha_{I+2})}\right]
    =
    -2iA^{(I+1)i}_{(\alpha_1,\ldots\alpha_{I+1})(\dot\alpha_1,\ldots\dot\alpha_{I+2})}
    +4i\sum_{m=1}^{I}\sum_{\substack{\{i\}\in\sigma^{I+1}\\\{j\}\in\sigma^{I+2}}}
    \binom{I}{m}\cdot\\\cdot
    \left[A^{(m-1)i}_{\alpha_{i_1},\ldots\alpha_{i_{m-1}},\dot\alpha_{j_1},\ldots\dot\alpha_{j_{m}}}
    ,C^{(I-m)}_{\alpha_{i_{m+1}},\ldots\alpha_{i_I},\dot\alpha_{j_{m+1}},\ldots\dot\alpha_{j_{I+2}}}\right]
    \eps_{\alpha_{i_m}\alpha_{i_{I+1}}}
\end{multline}
The $\sigma^n$ are all permutations of the integers $1,2,\ldots n$.
Recall that indices of $A$, $B$, $C$, $E$ and $G$ operators are
complectly symmetrized.

A key role in the next section is played by the first term in each
rhs. This term comes from the commutator
\begin{equation}\label{derv-under-charge}
    \Bigl[Q^i_{\alpha}\,,\,D_{\beta\dot\beta}\Bigr]\cdot\Op=
    -2\eps_{\alpha\beta}\Bigl[\bar\l_{\dot\beta}^i\,,\,\Op\Bigr]
\end{equation}
where $\Op$ is in the adjoint representation of $\SU(N)$.

\section{Fermi Surface Model of the Black Hole}\label{z-tit-fermi}

The model we propose for the operators corresponding to \bps{16}
\ads5 black hole microstates in the limit \eqref{eq:uslimit} is
based upon a fermi sea. Each fermion carries a fixed $SU(4)$ index
and an increasing angular momentum\footnote{i.e, the angular
momentum is analogous to the momentum for standard fermi surfaces.
Since we are working in radial quantization, or conversely, local
operators, this is a natural modification.}. In particular, the
difference between the two functional relations in \eqref{JbarJtoQ
ratio} and \eqref{JtoQ_ratio} comes about by the different ways of
filling the fermi-surface :  either by using $SU(2)_R$ singlets or
highest weight vectors.

Our fermi sea is constructed out of operators of the type
$A^{(I)\,1}$, as defined in equation \eqref{A-op-def}. To motivate
this, consider a black hole with $J={\bar J}$ (approximately),
satisfying $J>>\Qp$. We would like to construct operators out of the
basic fields in table-\ref{tab:fcharge}, having a large angular
momentum to R-charge ratio. The following restrictions apply:
\begin{itemize}
    \item We may use as many derivatives $D_{\alpha\dot\alpha}$ as
    needed.\vspace{-.4em}
    \item The BPS formula prevents us from using $\bar F_{\dot\alpha\dot\beta}$.\vspace{-.4em}
    \item The $M_{ij}$'s do not carry angular momentum and can be
    neglected at this stage of the construction.\vspace{-.4em}
    \item Fermionic operators $\l_{\alpha i}$ and $\bar\l_{\dot\alpha}^{i}$
    carry both angular momentum and R-charge. The BPS formula
    does not allow contractions of the $\SU(4)$ indices, thus the operators
    contribute only a linear relation between angular momentum and
    R-charge.\vspace{-.4em}
    \item $F_{\alpha\beta}$ carries no $\bar J$ and can be neglected when constructing an operator
    with $\bar J=J$.\vspace{-.4em}
\end{itemize}
This implies that the operator is made out of mainly gauge covariant
derivatives $D$ that increase the angular momentum ($J$) of a set
(order $\Qp$) of fields carrying the R-charge. Equation
\eqref{derv-under-charge} tells us that acting with the supercharges
$Q^i$ on any operator built out of many $D$'s, necessarily yields a
non-zero operator. A way to overcome this conclusion is to realize
that the "universal" part of the rhs side in
\eqref{derv-under-charge}  is a fermion - i.e, $\bar\lambda^i$.
Thus, if this fermion already appears in the operator, the Pauli
exclusion principle ensures that the descendant under $Q$ vanishes.

Two important properties of the $Q^1$ supercharges are :
\begin{subequations}
 \begin{align}
    \Bigl[Q^1, \underbrace{DD\cdots D}_{\text{$I$ times}}\cdot\Op\Bigr\}
    =&
    -2\sum_{m=1}^{I}\binom{I}{m}\Bigl[A^{(m-1)1},\underbrace{D\cdots D}_{\text{$(I-m)$
    times}}\cdot\Op\Bigr\}
    +\underbrace{DD\cdots D}_{\text{$I$ times}}\cdot\Bigl[Q^i,\Op\Bigr\}\,,
\\
    \Bigr\{Q^1,A^{(I)1}\Bigl\}=&-2\sum_{m=1}^{I}\binom{I}{m}\Bigl\{A^{(m-1)1},A^{(I-m)1}\Bigr\}\,.
    \label{eq:prop1}
\end{align}
\end{subequations}
Notice that all fermions appearing in the variation of the $D$'s are
always of type $A^{(I)\,1}$. Furthermore, the latter operators are
closed under the action of $Q^1$. We conclude that the simplest way
to make all the rhs supersymmetry variations of the $D$'s to vanish
is to use the fermions $A^{(I)\,1}$ as the basis for the fermionic
shells. The levels of these shells will be naturally labeled by the
left-handed angular momentum of the $A^{(I)\,1}$'s.

This simple assumption on the structure of the operators allows us
to reproduce the scalings  \eqref{JtoQ_ratio} and \eqref{JbarJtoQ
ratio} between R-charge and angular momentum up to coefficients of
order 1. Define the $\SU(2)_L\times\SU(2)_R$ highest weight
operator:
\begin{equation}\label{eq:highest weight}
    A^{(I)1}_{hw}\equiv A^{(I)1}_{\underbrace{{}_{111\ldots1}}_{I\text{
    times}},\underbrace{{}_{\dot1\dot1\dot1\ldots\dot1}}_{I+1\text{
    times}}}
\end{equation}
We focus on two cases :  ${\bar J}=0$ and ${\bar J}=J$.
\paragraph{The case $\bar J=0$ :}
The $\SU(2)_R$ invariant \bps{16} operators are built out of  a 'closed
fermi-surface' model (see figure-\ref{fig:closed}) described by the
operator
\begin{equation}\label{closed shell surface}
    \Jdp^{(K)}_{\text{closed}}\equiv\prod_{I=0}^{K}\prod_{m=0}^{I+1}\mathrm{Jdet}\left[(\bar J_-)^m A^{(I)\,1}_{hw}\right]
\end{equation}
where '$\mathrm{Jdet}$' stands for the anti-symmetrized
multiplication of the entire $\SU(N)$ adjoint
multiplet\footnote{This operation takes the vector space of the
adjoint into $(V)^{\wedge\dim G}$ which is a singlet.}:
\begin{equation}
    \mathrm{Jdet}\left[X\right]=\varepsilon_{a_1a_2\ldots a_{g}}X^{a_1}X^{a_2}\ldots
    X^{a_{g}}\csp1
    X=\sum_{a=1}^{g}X^aT^a\quad
    \left(g=\dim G\right)\,,
\end{equation}
for a fermionic X.

In a covariant form of \eqref{closed shell surface} the left-handed
angular momentum indices are totally symmetrized, whereas the action
of $\bar J_-$ generates the multiplication of the entire $\SU(2)_R$
multiplet. This causes $\bar J$ to vanish. Thus,
$\Jdp^{(K)}_{\text{closed}}$ belongs to the representation
$[k,0,0]_{(J,0)}$ and carries two charges $(J,\,\Qp)$ which for
large K and N equal:
\begin{align}\label{closed shell-charge}
    J=&(N^2-1)\sum_{I=0}^K\sum_{m=0}^{I+1}\frac{I}{2}=N^2\frac{K^3}{6}+O(K^2,N)\,,\cr
    \Qp=k=&(N^2-1)\sum_{I=0}^K\sum_{m=0}^{I+1}1=N^2\frac{K^2}{2}+O(K,N)\,.
\end{align}
Solving for $K$, the scaling $J\sim \Qp^{3/2}/N$ emerges, matching
\eqref{JtoQ_ratio}.

The operator $\Jdp^{(K)}_{\text{closed}}$ is invariant under the chiral supercharge
\begin{equation}\label{Jdp-supercharge}
    [k,0,0]_{(J,0)}\xrightarrow{~Q~}[k+1,0,0]_{(J\pm\frac12,0)}\\
    \Longleftrightarrow\quad\left[Q^{(i}_{\alpha},\Jdp^{(K)i_1,\ldots
    i_k)}_{\text{closed}}\right\}=0.
\end{equation}
This originates from the Pauli exclusion principle as follows. The
action of the supercharge 'splits' each $A$ factor in
$\Jdp^{(K)}_{\text{closed}}$ into two $A$ factors of smaller angular
momentum (see \eqref{eq:prop1}). However, each of these factors
already appears in $\Jdp^{(K)}_{\text{closed}}$. Thus
$\big[Q^1_\alpha,\Jdp^{(K)}_{\text{closed}}\big\}$ vanishes due to
its fermionic nature. Using the highest-weight of $\SU(2)_L$ and
$\SU(4)$ the construction can be viewed as a fermionic shell model,
whose 'level' is the left-handed angular
momentum\footnote{Remembering that for the $A^{(I)1}$ building
blocks $\bar J=J+1/2$, thus we can use the right-handed angular
momentum as well.} and the degeneracy is the $\SU(2)_R$ and $\SU(N)$
multiplet. In this picture, each $A$ factor is a creation operator
of a fermionic state, and consequently, the
$\Jdp^{(K)}_{\text{closed}}$ corresponds to filling all the shell up
to level K. In terms of figure \ref{fig:closed}, for each level
$\bar J$ (equally $J$) we fill all the $\SU(2)_R$ multiplet $\bar
J^3=-\bar J,\ldots \bar J$. The action of the chiral supercharge
tries to split a fermion into two fermions belonging to lower
levels, which is forbidden due to Pauli exclusion.


\paragraph{The case $\bar J=J$ :}
The equal left and right angular momenta \bps{16} operators are
built out of an 'open fermi-surface' model (see figure-\ref{fig:open})
described by the operator
\begin{equation}\label{open shell-surface}
    \Jdp^{(K)}_{\text{open}} \equiv\prod_{I=0}^{K}\mathrm{Jdet}\left[A^{(I)\,1}_{hw}\right]
\end{equation}
The absence of $\bar J_-$ (compared to the closed shell model) in a
covariant form causes all Lorentz indices (left and right handed) to
be fully symmetrized. In terms of figure \ref{fig:open} at each
level $\bar J$ (equally $J$) we occupy a single fermion with maximal
$\bar J^3=\bar J$ . Calculating the charges in the large K and large
N regime:
\begin{align}\label{open shell-charge}
    J=\bar J=&(N^2-1)\sum_{I=0}^K\frac{I}{2}=N^2\frac{K^2}{4}+O(K^2,N)\,,\cr
    \Qp=k=&(N^2-1)\sum_{I=0}^K1=N^2K+O(K,N)\,.
\end{align}
Once again, solving for $K$, the scaling $J=\bar J\sim \Qp^{2}/N^2$
emerges, matching \eqref{JbarJtoQ ratio}.
\DOUBLEFIGURE{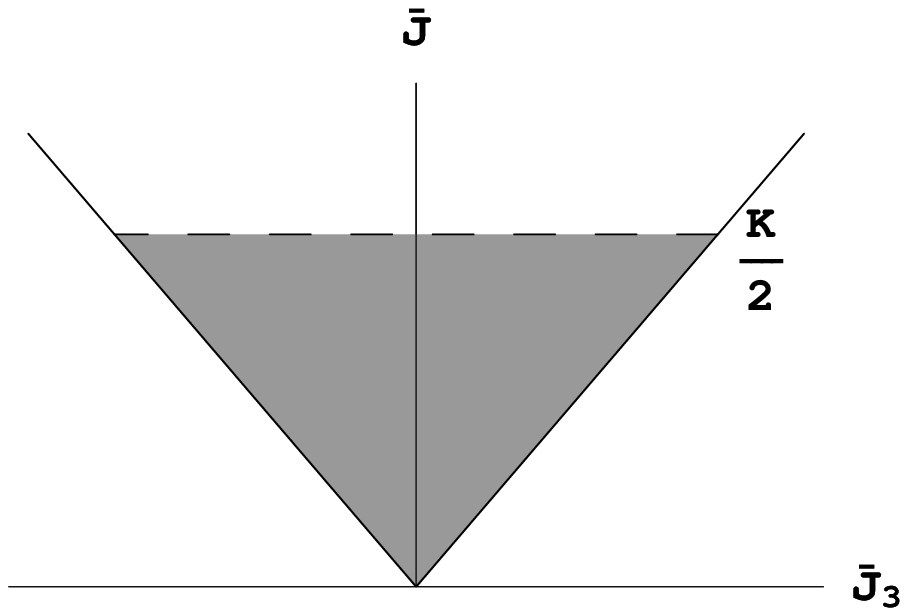,width=6.5cm}{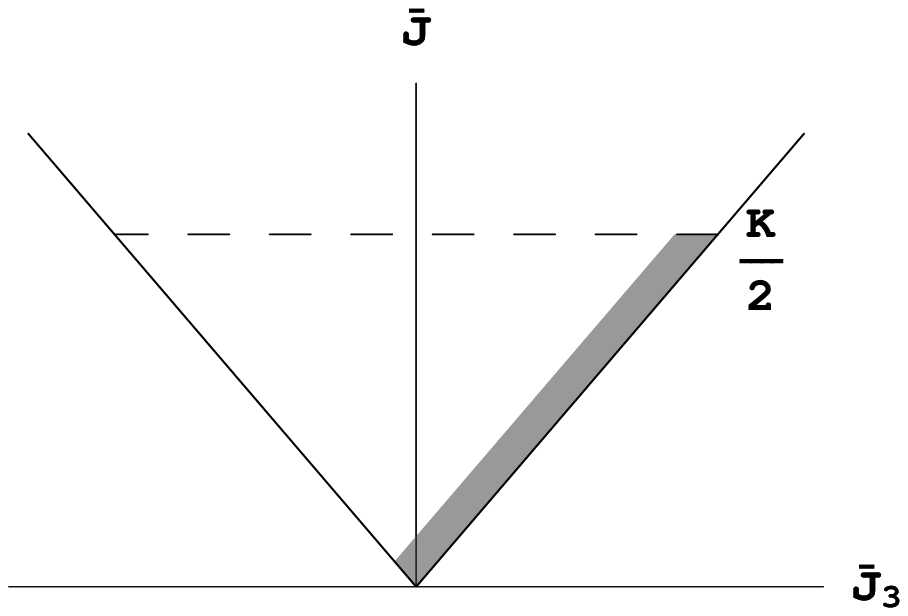,width=6.5cm}{\label{fig:closed}The
fermi sea picture of the close-shell operator}{\label{fig:open}The
fermi sea picture of the open-shell operator}

\paragraph{}
The operators introduced for the fermi-surface models are manifestly
descendants, as seen from the "extra" supercharge in
\eqref{Jdp-supercharge} and the failure to satisfy the BPS
formula\footnote{One may wonder if the failure to comply with the
BPS bound should means that the operator vanish, it is easy to check
that this is not the case for the case of $\SU(2)$ where one can
replace the determinant of the adjoint representation by a trace
$\mathrm{Jdet}\left[X\right] =
\tr_{(\text{fund})}\Big(X\{X,X\}\Big)$ for a fermionic X}:
\begin{equation}
    \Delta_{BPS}[k,0,0]_{(J,0)}=2+2J+\frac32k\qquad
    \Delta[\Jdp^{(K)}] = 2J+\frac32k
\end{equation}
In the rest of the section we show how to construct genuine \bps{16}
primaries by combining the fermi sea with bosonic operators. It is
the addition of these bosonic excitations that yields a macroscopic
entropy, i.e large enough degeneracy of operators, to generate a
macroscopic black hole entropy in Planck units. There is also a
large degeneracy of fermi surfaces as we will see in section
\ref{z-tit-gen}.

\subsection{Building $c^{1/4}$ Primaries}\label{z-tit-buil}

We are interested in modifying the shell construction to achieve
several goals :  saturation of the BPS bound, introduction of
degeneracies (entropy) and having the operator be a primary. All
these properties are satisfied by the addition of the adequate
bosonic structures. In particular, we consider the following large
family of $c^{1/4}$ operators :
\begin{equation}\label{Closed shell}
    \Op^{(K,\vec J)}\equiv \Jdp^{(K)}\left(C^{(K+1)}\Bp^{(\vec
    J)}\right)_{GI}
    \qquad 2J_i\leq K
\end{equation}
$\Jdp^{(K)}$ stands for either \eqref{closed shell surface} or
\eqref{open shell-surface} and the subscript 'GI' stands for a gauge
invariant combination. $\vec J$ is a length $3L$ vector of angular
momenta and
\begin{equation}\label{primay-op}
    \Bp^{(\vec J)}\equiv\prod_{i=1}^{L}B^{(2J_i)\;12}B^{(2J_{L+i})\;13}B^{(2J_{2L+i})\;14}\,.
\end{equation}
Notice that we are forced to add the bosons $B^{(I)1i}$ in triplets
to have vanishing R-charges $p$ and $q$. The operator $C^{(K+1)}$ is
the only building block in $\Op^{(K,\vec J)}$ satisfying
$\Delta_{exc}=\Delta-2J-\frac32k=2$ (all the rest has
$\Delta_{exc}=0$). This suggests including a single excitation of
type $C$ in each $c^{1/4}$ operator to saturate the BPS bound. As we
explain below, the insertion of $C$ also plays a crucial role in
allowing the full operator \eqref{Closed shell} to be a primary.

For the closed shells models the right-handed angular momentum
coming from the $C^{(K+1)}$ and $\Bp^{(\vec J)}$ is arbitrary. For
the open shells model we need to symmetrize over all doted indices
coming from the fermion and bosons resulting in $\bar J\simeq J$.
Actually for the open shells one needs to work a little harder to create
a primary. The total $\bar J$ is larger than $J$ due to the extra
doted index of the Weyl spinors $\bar \l^1_{\dot\alpha}$, with the
consequence that we are really describing a descendant operator. In
section \ref{z-tit-hybr} we show how to fix this problem.

Acting with the chiral-supercharge \eqref{Jdp-supercharge} on the
bosons $B^{(I)1i}$ and $C^{(K+1)}$ splits any boson into a sum of
pairs consisting of a fermion and a boson in lower levels. All the
fermions are of the type $A^{(I)1}$. Our operators are constructed in
such a way that all $A^{(I)1}$ operators generated from the splitting of
$B$ or $C$ are occupied. Hence $Q^1$ acting on $B$ or $C$ vanishes on
the fermi surface (this is the origin of the constraints on the
maximal $J_i$'s in \eqref{primay-op}). The only exception is the
supercharge acting on the $C^{(K+1)}$ which contains a term
transforming it to a fermion in a higher level. Thus we are left
with:
\begin{equation}
    \left[Q^{(i}_{\alpha},\Op^{(K,\bar J)i_1,\ldots i_k)}\right\}=
    \Jdp^{(K)}\left(A^{(K+2)1}\Bp^{(\vec J)}\right).
\end{equation}
In term of charges
\begin{equation*}
    [k,0,0]_{(J,\bar J)}\xrightarrow{~Q^1~}[k+1,0,0]_{(J+\frac12,\bar J)}
\end{equation*}
The above argument proves that $\Op^{(K,\bar J)}$ obeys the
semi-shortening condition of a super-multiplet with
$[k+1,0,0]_{(J-\frac12,\bar J)}$ removed. We are still left with the
task of finding out when the operator $\Op^{(K,\bar J)}$ is a
primary.

For the $\vec J=0$ case, we would like to suggest the following
criteria for the bosonic part of the operator (although a full proof
remains to be carried out). The constraint for $\Op^{(K,\bar 0)}$ to
be a primary\footnote{up to the addition of descendants, of course.}
is that its bosonic part (i.e, its $B$'s + a single $C$) is an
$\N=4$ \bps8 operator, with the only difference being that a single
$C^{(K+1)}$ is plugged into one of the traces.

The arguments for this claim are the following. The composite
$\Op^{(K,\vec 0)}$ is made out of three components:
\begin{enumerate}
    \item The $B$'s part is a genuine \bps8 operator, annihilated only by
    $Q^{1}_{\alpha}$, and cannot be written as a $Q$, $\bar Q$ or a
    derivative of anything.
    \item The $C^{(K+1)}$ part, which can be written as (no summation of repeated indices):
    \begin{equation*}
        C^{(K+1)}=\left(\{Q^1,\bar Q_1\}\right)^{K+1}\cdot\{\bar Q_i,[\bar Q_j\bar M^{ij}]\}
    \end{equation*}
    \item The closed shells operator $\Jdp^{(K)}$
\end{enumerate}
We argue that any attempt to write $\Op^{(K,\vec 0)}$ as a $Q$ or a
$\bar Q$ of another operator, just by "pulling out" a single
supercharge fails. Our arguments are not complete, but we analyse
the simplest ways to write $\Op^{(K,\vec J)}$ as a $Q$ or $\bar Q$
of another operator.

First we try "pulling out" a supercharge from one of the components.
We cannot pull out anything from the $B$'s part, so we try to write:
\begin{equation}\label{primary_proof1}
    \Op^{(K,\vec 0)}=Q\cdot\left(\Jdp^{(K)}Y \Bp^{(\vec 0)}\right)
    \qquad\text{or}\qquad
    \Op^{(K,\vec 0)}=Q\cdot\left(X C^{(K+1)}\Bp^{(\vec 0)}\right)
\end{equation}
For the above to "work" we need $Q$ to annihilate $\Bp^{(\vec 0)}$,
 the only possible supercharges are $Q^1_{\alpha}$. Considering
the supersymmetry transformation, we see that $X$ and $Y$ in
\eqref{primary_proof1} are\footnote{The notations $Q^1_{+1/2}$ and
$Q^1_{-1/2}$ stand for the parts of the supercharge which raise or
lower the angular momentum (respectively).}:
\begin{align*}
    C^{(K+1)}=&\,Q^1_{+1/2}\cdot A^{(K)1}
    &\Rightarrow&& Y=&\,A^{(K)1}\\
    \Jdp^{(K)}=&\,Q^1_{-1/2}\cdot\pdf{^2\Jdp^{(K)}}{A^{(I)1}A^{(I')1}}A^{(I+I'+1)1}
    &\Rightarrow&& X=&\,\pdf{^2\Jdp^{(K)}}{A^{(I)1}A^{(I')1}}A^{(I+I'+1)1}.
\end{align*}
In the above expression, a derivative of a composite with respect to fermionic
operators should be understood as removing a single copy of the
operator from the composite.

The first option fails, due to Pauli exclusion - the operator Y is
annihilated on the fermi sea. The second option inserts holes in
the fermi sea at $A^{(I)1}$ and $A^{(I')1}$. This means that the
variation has extra terms coming from the variation of $C^{(K+1)}$
which fills one of these holes. Hence, we do not obtain equation
\eqref{primary_proof1} in this way. It does not seem possible to
cancel these extra terms (for example, by taking sums over different
$I$ and $I'$), although a full proof remains to be formulated.

The next possibility we attempt to falsify is "pulling out" a
supercharge from the combination of the fermi-sea and bosons
$\Bp^{(\vec 0)}$, i.e splitting a boson into a pair of a fermion and
a boson:
\begin{equation*}
    \Op^{(K,\vec 0)}=Q\cdot\left(\pdf{\Jdp^{(K)}}{A^{(I)1}}C^{(K+1)}\pdf{\Bp^{(\vec 0)}}{B^{(0)1i}}B^{(I+1){i'j'}}\right)
\end{equation*}
Checking the supersymmetry transformations, we see that the only
possibility is having $i'j'=1i$ and $Q=Q^{1}_{-1/2}$. Now we can
repeat the argument that the hole in the fermi-sea allows for
non-vanishing transformation of the $C^{(K+1)}$ and fails to achieve the
above equality.

Trying to "pull out" a supercharge form the combination of the
$C^{(K+1)}$ and $\Bp^{(\vec 0)}$, fails from similar reasonings. We
are left to check that we cannot "pull out" a supercharge from the
combination of the fermi-sea and the $C^{(K+1)}$. To examine this
option, consider the supersymmetry transformation:
\begin{multline}\label{desc-atm}
    Q^1_{-\frac12}\cdot\left(\pdf{\Jdp^{(K)}}{A^{(I)1}}C^{(J+I+1)}\Bp^{\vec(0)}\right)=\\
    =\Jdp^{(K)}\,C^{(J)}\Bp^{(\vec0)}
    +\sum_{r=K+2}^{J+I+1}\pdf{\Jdp^{(K)}}{A^{(I)1}}A^{(r-1)1}C^{(J+I+1-r)}\Bp^{(\vec0)}
\end{multline}
We would like to know for what values of $J$, the sum in the rhs
will be non-zero for any value of I ($I\leq K$). In addition for
$\Jdp^{(K)}\,C^{(J)}\Bp^{(\vec0)}$ to be BPS we must have ~$J<K+2$.
The conditions that $J<K+2$ and that $J+I+1\geq K+2$ for all $0\leq
I\leq K$ have a unique solution of $J=K+1$, which is the operator
that we presented before.

From the above discussion we also learn the existence of a general rule: in order
to construct a primary from a fermionic shell model, we must have a
$C$ factor in an empty shell adjacent to the last filled shell.

\subsection{Charges}\label{z-tit-charges}

$\Op^{(K,\bar J)}_{\text{closed}}$ and $\Op^{(K,\bar
J)}_{\text{open}}$  are constructed so that the $\SU(2)_L$ and
$\SU(4)$ charges are additive. The contributions of each composite
to the global charges\footnote{In this section we are explicitly
using the Cartan of the the $\SU(4)$ R-symmetry, the Dynkin labels
$(k,p,q)$ are the weights of the states.} carried by the operator
are summarized in table-\ref{tab:hw-charges}.\TABLE{
\begin{tabular}{|c|c|c|c|c|} \hline
              & ~~~$J$~~~& $~~~(k,p,q)~~~$ &~~~$\Delta$~~~&$~~\Delta-2J-\frac32 k$~~\\
\hline
 $A^{(I),1}$  &  $I/2$   &   $(1,0,0)$  & $I+3/2$  & $0$       \\
 $B^{(I)12}$  &  $I/2$   &   $(0,1,0)$   & $I+1$    & $1$      \\
 $B^{(I)13}$  &  $I/2$   &   $(1,-1,1)$  & $I+1$    & $-1/2$   \\
 $B^{(I)14}$  &  $I/2$   &   $(1,0,-1)$  & $I+1$    & $-1/2$   \\
  $C^{(I)}$   &  $I/2$   &   $(0,0,0)$  & $I+2$    & $2$       \\
\hline
\end{tabular}
\label{tab:hw-charges}\caption{Charges of $\Jdp^{(K,\vec J)}$
building blocks.}}

Remembering that the $B$'s come in triplets $B^{12}B^{13}B^{14}$,
we immediately see the emergence of the BPS formula:
\begin{equation}
    \Delta=2+2J+\frac32k\,.
\end{equation}

We calculate the charges of the closed shell model with $B$'s,
postponing the open shell model discussion to section
\ref{z-tit-hybr}. For the closed shells model, the total charges are
computed by summing the contributions over the different
ingredients:
\begin{subequations}
 \begin{align}
    J=&(N^2-1)\sum_{I=0}^K\sum_{m=0}^{I+1}\frac{I}{2}+\frac{K+1}{2}+\sum_{i=1}^{3L}J_i\\
    \Qp=k=&(N^2-1)\sum_{I=0}^K\sum_{m=0}^{I+1}1+2L
\end{align}
\end{subequations}
The right handed angular momentum is bounded from above by
$\sum_{i=1}^{3L}J_i$, but could be taken to $0$  by suitable
contractions.

Taking the large R-charge and large angular momentum limit is
equivalent to taking $K\gg1$. Simplifying the charges in this case
and taking $N\gg1$:
\begin{subequations}
 \begin{align}
    &J\approx N^2\frac{K^3}{6}+\sum_{i=1}^{3L}J_i\leq N^2\frac{K^3}{6}+\frac32L K\,,\label{J-bound1}\\
    &\Qp\approx
    N^2\frac{K^2}{2}+2L\,,\label{Q-charge1}
\end{align}
\end{subequations}
where the bound in \eqref{J-bound1} originates from \eqref{Closed
shell}. The maximal value for $J$ in this family of operators is obtained
as follows. First, we solve \eqref{Q-charge1} for L and substitute
it back into \eqref{J-bound1}
\begin{equation}
    J\leq \frac{3\Qp}4K-\frac{5}{24}N^2 K^3\,.
\end{equation}
If we view the rhs as a function of K, the latter is bounded from above. This generates an upper bound
for $J$ for all $\Qp$ given by
\begin{equation}
    \frac{J_{max}}{N^2}=\frac{\sqrt{3}}{\sqrt{20}}{\sqrt2}\left(\frac{\Qp}{N^2}\right)^{3/2}
    \approx0.39\,{\sqrt2}\left(\frac{\Qp}{N^2}\right)^{3/2}
\end{equation}

If we compare this result to the supergravity scaling
\eqref{JtoQ_ratio} (with $\gamma=0$), we realize that our fermi-sea
operators reproduce the same scaling relation, but differ in an
order one number in its coefficient. In particular, the angular
momentum is approximately $0.39$ times smaller than the supergravity
charge. If we had neglected the bosons,  we would have found :
\begin{equation*}
    \frac{J}{N^2}=\frac{\sqrt2}{3}\left(\frac{\Qp}{N^2}\right)^{3/2}
\end{equation*}
Thus, the addition of the bosons improves the order 1 coefficient
but not enough to match the supergravity result.

One can also wonder about lower values of $J$. Naively one can add
many bosons in low angular momentum levels. Such operators have
angular momentum linear in the charge (or less). For example, adding
bosons up to level $K'$ which is fixed as $K$ scales to infinity,
pulls down the angular momentum to charge scaling down to
$J\sim\Qp^{\beta}$ ($0\leq\beta<3/2$). However, the degeneracy of
such configuration is of the same order as the degeneracy of
standard \bps8 operators which scales as $N\log N$
\cite{Kinney:2005ej}. As we discuss in the following section, the
entropy of our operators, with large angular momentum ($J\sim
\Qp^{3/2}/N$), scales as $\Qp$ (which is much greater than $\gg
N^2$). Thus operators with scaling $\beta<3/2$ are subdominant and
should not affect the macroscopical features of the ensemble.

The shell structure that we discussed, and its completion to primary
operators, reproduces the scaling relation $J(\Qp)$ up to numerical
coefficient. We now present a simple computation that reproduces the
correct scaling of the entropy as well, up to order 1 coefficient.
We carry out the computation both for the open and closed shells. In
both cases, the entropy will be proportional to $\Qp$, which is the
correct result, but we will see that it comes about in different
ways for the two cases.

\subsubsection{Entropy of the Closed Shell Model}

In this section we estimate the degeneracy of the bosonic part under
the following assumptions :
\begin{itemize}
    \item Ignoring the constraints for the operator to be primary.
    \item Ignoring any finite $N$ dependence.
\end{itemize}
The statistical model we use is a Fock space of free bosons. The
single particle bosonic states contribution to the degeneracy are:
$B^{(I)12}$, $B^{(I)13}$ and $B^{(I)14}$, with I taking values from
$0$ to $K$. We introduce a chemical potential for the
right-handed angular momentum ($J$) and for the R-charges ($k$, $p$
and $q$) allowing for $\bar J$ to be determined by the ensemble
average. The partition function takes the familiar form (see
\cite{Aharony:2003sx}) of summation over all multi-particle states:
\begin{equation}
    \log Z=\sum_{r=1}^{\infty}\frac1{r}f_{sp}(r\gamma_1\,,\,r\gamma_2\,,\,r\gamma_3\,,\,r\mu)\csp1
\end{equation}
with the single particle partition function:
\begin{equation}
    f_{sp}(\gamma_1\,,\,\gamma_2\,,\,\gamma_3\,,\,\mu)=
    \sum_{I=0}^K\sum_{a=1}^{N^2-1}\sum_{m=0}^{I}
    \left(e^{\gamma_1-\gamma_3}+e^{\gamma_1-\gamma_2+\gamma_3}+e^{\gamma_2}\right)e^{\frac{\mu I}{2}}
\end{equation}
The chemical potentials are defined such that the boson contribution
to the charges is :
\begin{equation}
    \hat k=\pdf{\log Z}{\gamma_1}\,,\qquad
    \hat p=\pdf{\log Z}{\gamma_2}\,,\qquad
    \hat q=\pdf{\log Z}{\gamma_3}\,,\qquad
    \hat J=\pdf{\log Z}{\mu}\,.
\end{equation}
We are interested in $p=q=0$, which determines :
\begin{equation}
    \gamma_1=3\gamma_3=\frac{3}{2}\gamma_2\equiv\gamma
\end{equation}
Evaluating the partition function (in the large $N$ and $K$ limit):
\begin{align}
    \log Z(\mu,\gamma)=&-3N^2\sum_{I=0}^{K}(I+1)\log\left[1-
    \exp\left(\frac23\gamma+\frac{\mu}{2}I\right)\right]=\cr
    \approx&-3N^2K^2\int_0^{1}dy\,y\log\left[1-
    \exp\left(\frac23\gamma+\frac{\mu}{2}yK\right)\right]=\cr
    =&\frac{12K^2N^2}{(K\mu)^2}\left(
    \pl{3}{e^{\frac23\gamma}}
    -\pl{3}{e^{\frac23\gamma+\frac{K\mu}{2}}}
   +\frac{K\mu}{2}\pl{2}{e^{\frac23\gamma+\frac{K\mu}{2}}}\right)\cr
\end{align}
where $\pl{n}{z}$ is the PolyLog function. The form of the partition
function suggests using the variables:
\begin{equation*}
    x=\frac{K\mu}{2}\csp1 \xi=e^{\frac23\gamma}
\end{equation*}
We wish to set the chemical potentials to fix the charges
(remembering the contribution of the fermions):
\begin{align}\label{ensembel-charge}
    J=&\frac{N^2K^3}{6}+\hat J=\frac{N^2K^3}{6}+\frac{K}{2}\pdf{\log
    Z}{x}=N^2K^3\,a(x,\xi)\cr
    \Qp=&\frac{N^2K^2}{2}+\hat k=\frac{N^2K^2}{2}+\frac23\pdf{\log Z}{\log\xi}
    =N^2K^2\,b(x,\xi)
\end{align}
with,
\begin{align}
    a(x,\xi)\equiv&\,\frac1{6}-\frac{3}{2x}\log\left(1-\xi e^x\right)
    -\frac{3}{x^3}\left(\pl{3}{\xi}-\pl{3}{\xi e^x}+x\pl{2}{\xi e^x}\right)\cr
    b(x,\xi)\equiv&\,\frac1{2}-\frac{2}{x}\log\left(1-\xi e^x\right)
    +\frac{2}{x^2}\left(\pl{2}{\xi}-\pl{3}{\xi e^x}\right)
\end{align}
There are two conditions that we would like to force on the
ensemble:
\begin{equation}\label{ensembel-cond}
    \frac{J}{N^2}={\alpha}{\sqrt2}\left(\frac{\Qp}{N^2}\right)^{3/2}
    \csp1
    J\gg\Qp
\end{equation}
The first is the supergravity scaling (from the previous discussion
we expect $\alpha$ to be of order 1). The second is the condition
for the energy to be dominated by the angular momentum. Applying
\eqref{ensembel-charge} to \eqref{ensembel-cond} we conclude:
\begin{equation*}
    a\sim b^{3/2}\csp1
    K\gg\frac{b}{a}\sim b^{-1/2}
\end{equation*}
The interesting regime is $K$ large, $a$,$b$ fixed ($x$, $\xi$
fixed). In this regime the scaling of the entropy becomes :
\begin{align}
    S(J,\Qp)=&\,\log Z-\frac23\gamma\cdot\hat k-\frac{\mu}{2}\cdot\hat J
    =K^2N^2f(x,\xi)\cr
    \sim&\,K^2N^2\sim\, \Qp
\end{align}
Since the variable $x$ and $\xi$ depend on $\Qp$ and $J$ in such a
manner that they do not scale with $K$ or $N$, therefore the
function $f(x,\xi)$ does not scale with $K$ or $N$.

This result matches qualitatively the supergravity relation
\eqref{entropy-sugra} and confirms our claim that the \bps{16}
operators constructed from the closed shell models indeed carry
macroscopical large entropy ($S\sim N^2K^2$) unlike the \bps8
operators with no angular momentum ($S\sim N\log N$).

In the above calculation we used a slightly different scheme than in
the rest of the paper. We fixed $J$ and $\Qp$ and let $\bar J$ be
determined by the ensemble (instead of fixing $\bar J$ and $\Qp$).
This was done for convenience and it should not affect the validity of our conclusion.

\subsubsection{Entropy of the Open Shells Models}

The calculation for the open shells is done in a similar spirit to
the closed shells one with the summation over the $\SU(2)_R$
multiplet removed.
\begin{equation}
    \log Z=\sum_{r=1}^{\infty}\frac1{r}f_{sp}(r\gamma\,,\,r\mu)\csp1
    f_{sp}(\gamma\,,\,\mu)=
    \sum_{I=0}^K\sum_{a=1}^{N^2-1}
    3e^{\frac23\gamma+\frac{\mu I}{2}}
\end{equation}
Repeating the steps of the previous subsection, we obtain :
\begin{align}\label{ensembel-charge-op}
    J=&\frac{N^2K^2}{4}+\hat J=N^2K^2\,a(x,\xi)\cr
    \Qp=&N^2K+\hat k=N^2K\,b(x,\xi)
\end{align}
with,
\begin{align}
    a(x,\xi)\equiv&\,\frac1{4}-\frac{3}{2x}\log\left(1-\xi e^x\right)
    -\frac{3}{2x^2}\left(\pl{2}{\xi e^x}-x\pl{2}{\xi}\right)\cr
    b(x,\xi)\equiv&\,1-\frac{2}{x}\left(\log\left(1-\xi
    e^x\right)-\log\left(1-\xi\right)\right)
\end{align}
The two conditions that we force on the ensemble are :
\begin{equation}\label{ensembel-cond-op}
    \frac{J}{N^2}=\alpha\left(\frac{\Qp}{N^2}\right)^{2}
    \csp1
    J\gg\Qp
    \rar1
    a\sim b^2\csp1
    K\gg b^{-1}
\end{equation}
We find that  $a(x,\xi)$ and $b(x,\xi)$, cannot scale with $K$ or
$N$. In general, solving \eqref{ensembel-charge-op} for $x$
and $\xi$, we conclude that all the dependence on $\Qp$ and $J$ is
such that they are order 1 numbers (not scaling with $K$ or $N$).
Thus, the entropy of the ensemble is given by :
\begin{align}
    S(J,\Qp)=&\,\log Z-\frac23\gamma\cdot\hat k-\frac{\mu}{2}\cdot\hat J
    =K^2N f(x,\xi)\cr
    \sim&\,K^2N\sim\, \Qp
\end{align}
The result matches qualitatively the supergravity relation
\eqref{entropy-sugra-jbar j} and matches our expectations that the
\bps{16} operators constructed from the open shell models indeed
carry macroscopical large entropy ($S\sim N^2K$).

\section{Generalizations}\label{z-tit-gen}

In this section, we take some steps towards generalizing the
structures studied before. Even though one can potentially achieve
this by adding more fields to the operators in question, we focus
here on a more interesting possibility which is the deformation of
the fermi sea structure. We will not describe the bosonic part of
the operator nor discuss whether these new operators are primary or
not.

First of all, we provide the basic rules that any shell has to
satisfy. A fermion in the fermi-sea is characterized by its quantum
numbers under  $\SU(2)_L\times SU(2)_R$, i.e. \footnote{We
temporarily change our notation, using the quantum numbers instead
of the Lorentz indices. We use $m$ and $\bar m$ for the $J^3$ and
$\bar J^3$ eigenvalues, respectively.}
~$A_{\ket{J,\,m=J}\ket{\bar{J}=J+1/2,\,\bar m}}$. Under the action
of the supercharge \eqref{supercharge}, the fermion splits according
to :
\begin{gather}
    A_{\ket{J,J}\ket{J+\frac12,\,\bar m}}\xrightarrow
    {Q^1_{-1/2}}
    \sum_{\substack{J_1,J_2\\\bar m_1,\bar m_2}}A_{\ket{J_1,J_1}\ket{J_1+\frac12,\,\bar m_1}}
    \otimes A_{\ket{J_2,J_2}\ket{J_2+\frac12,\,\bar  m_2}}
\cr
    \text{with,}\qquad  J=J_1+J_2+1/2\csp1\bar m=\bar m_1+\bar m_2
    \label{spliting}
\end{gather}
Denoting the set of occupied fermions by ${\cal M}$, the conditions
for invariance under \eqref{supercharge} are that
~$A_{\ket{J_1,J_1}\ket{J_1+\frac12,\,\bar m_1}}\in\mathcal{M}$
or,~$A_{\ket{J_2,J_2}\ket{J_2+\frac12,\,\bar m_2}}\in\mathcal M$ for
each possible combination in the sum \eqref{spliting}.

Figures \ref{fig:band} and \ref{fig:hybrid} exhibit two methods of
finding a set ${\cal M}$ satisfying these constraints (there are
also ways of combining the two methods). In section
\ref{z-tit-ferm-bands} we discuss the method corresponding to
\ref{fig:band}, and in section \ref{z-tit-hybr} we discuss the
generalization corresponding to \ref{fig:hybrid}.

\DOUBLEFIGURE{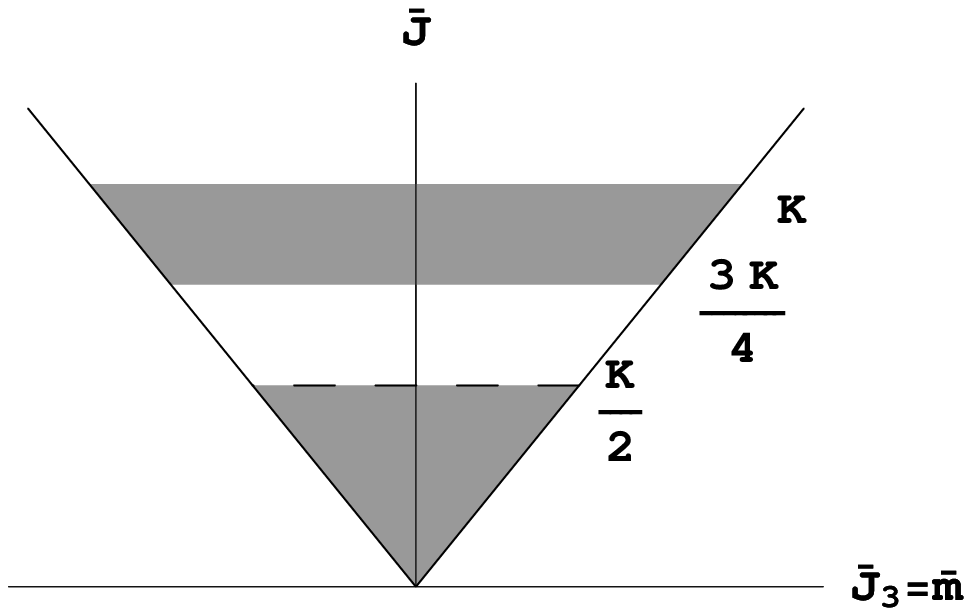,width=6.5cm}{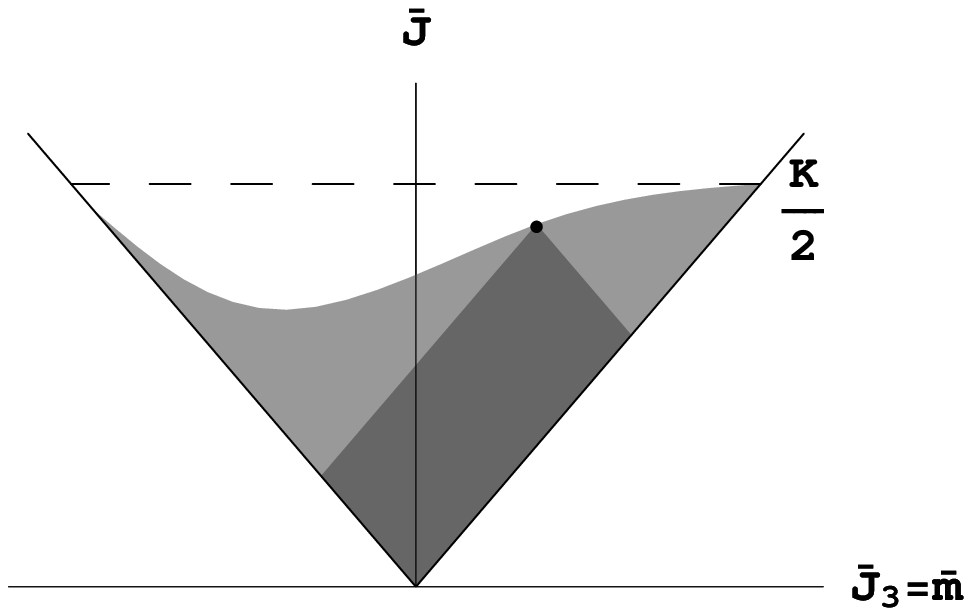,width=6.5cm}{\label{fig:band}A
band of closed-shells}{\label{fig:hybrid}A general fermi sea
picture}

\subsection{Fermionic Bands}\label{z-tit-ferm-bands}

The first generalization that we describe is to add fermions in a
level higher than $K$. Any fermionic operator $A^{(I)\,1}$ with
level up to $2K$ splits under the supercharge $Q^1$ action into two
fermions, such that at least one of them is below level $K$. Hence,
such action is annihilated on the closed shell.  We can continue
this construction by adding closed shells near level $2K$ (we call
this a band) allowing to have fermions with level up to $3K$.

Iterating this procedure, we can build multiple fermionic bands.
Leaving the details to appendix-\ref{z-tit-apd-fermion}, we search
for the best configuration of fermions in $n$ bands. The upper bound
on the angular momentum to charge ratio is found for the single-band
case drawn in figure-\ref{fig:band}, with:
\begin{equation}
    J=\alpha{N^2}{\sqrt2}\left(\frac{\Qp}{N^2}\right)^{3/2}\csp2
    \alpha\leq\frac{3}{2\sqrt{11}}\approx0.45.
\end{equation}
The bound is saturated when the contribution of the bosons is
completely negligible\footnote{In practice, we need a small number
of bosons to satisfy the primary conditions.}.

%
%
%
\subsection{General Fermionic Shells}\label{z-tit-hybr}

As with any fermi surface, we can deform it. For our surface in the
${\bar J}-{\bar m}$ plane, this can be done as follows. Regarding
$(\bar J,\bar m)$ as a 2-vector, we see that the splitting of a
fermion in \eqref{spliting} by the supercharges results in a
2-vector summation:
\begin{equation*}
    (\bar J,\bar m)=(\bar J_1,\bar m_1)+(\bar J_2,\bar m_2)
    \csp2
    \bar J_i\geq\abs{\bar m_i}\quad (i=1,2)\,,
\end{equation*}
where the last inequality is the condition that the vector
represents true $\SU(2)_R$ quantum numbers. Therefore a fermion can
only split into parts that are confined to a rectangular whose
opposite corners are the original vector and the origin (described
in figure \ref{fig:hybrid} by the darker part of the fermi-sea).

Hence, the description of the fermi-sea is given by the contour of
the last (highest angular momentum) occupied fermions $\bar
J_{max}(\bar m)$. The condition for invariance under the
supercharge is simply: 
\begin{equation*}
    \abs{\frac{d\bar J_{max}}{dm}}\leq1
\end{equation*}
Even though the calculation of charges of the surface is somewhat complicated,
the value of $\Qp$ , $J$ and $\bar J^3$ are just integrals (in the
large $K$ limit) over the fermi-sea:
\begin{subequations}
 \begin{align}
    \frac{J}{N^2}=&\iint_{sea}d\bar Jd\bar m\,\bar J\label{J-integral}\\
    \frac{\Qp}{N^2}=&\iint_{sea}d\bar Jd\bar m\,1\\
    \frac{\bar J^3}{N^2}=&\iint_{sea}d\bar Jd\bar m\,\bar m
\end{align}
\end{subequations}
For \eqref{J-integral} recall that we have chosen a highest weight
with respect to $\SU(2)_L$ (see eq. \ref{eq:highest weight}). A
state constructed this way has a well defined $\bar J^3$ eigenvalue,
but one still needs to project to states with specific $\bar J$.

In the following paragraph we describe in detail an example for a
class of fermi surfaces where we have a good control over all
charges. The operators in this class have the nice feature that they
have scalings matching the black holes with arbitrary $J$ and $\bar
J$.

\subsubsection{Generalized open-shells}\label{z-tit-constant}

The fermi sea we will describe is a generalization of the open-shell
model, where a constant number of fermions at each level is kept.
The corresponding fermi-sea is drawn in figure-\ref{fig:g-opens}.
\FIGURE{\centerline{\vbox{\hspace{3cm}\includegraphics[width=7cm]{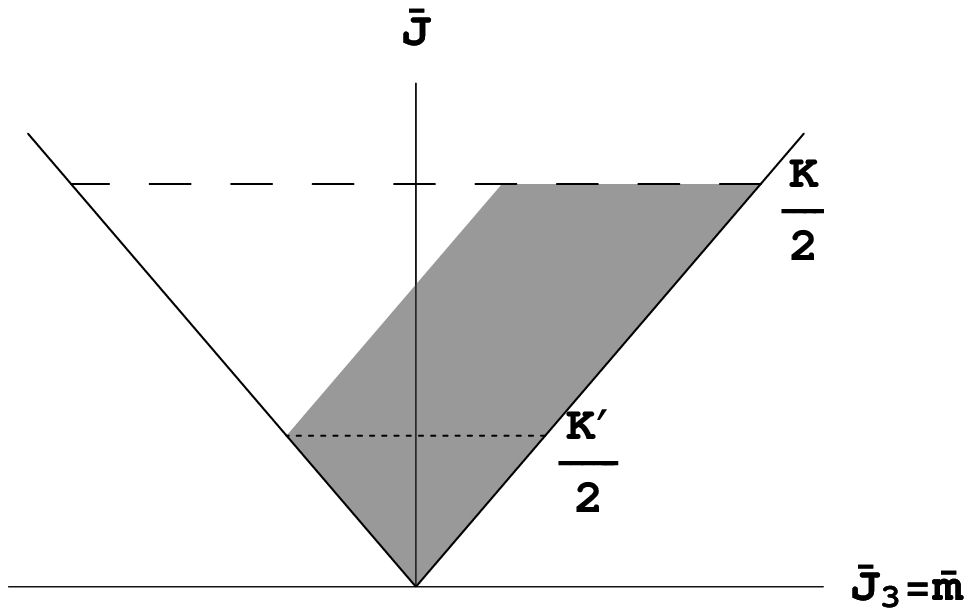}
\caption{Fermi sea of generalized open-shells
\label{fig:g-opens}}}}}

In order to construct these operators, we need to be more explicit with the $\SU(2)_R$ symmetry. First, rewrite
the closed shell operator in a manifestly covariant form:
\begin{equation}
    \Jdp^{(K)}_{\text{closed}}\equiv\prod_{I=0}^{K}\prod_{m=0}^{I+1}\Jdet\left[(\bar J_-)^m A^{(I)1}\right]
    =\prod_{I=0}^{K}\Jdet\left[\Adp^{(I,I+1)}\right]
\end{equation}
where,
\begin{align}
    \Adp^{(I,I+1)}=\prod_{m=0}^{I+1}(\bar J_-)^m A^{(I)1}
    \equiv&~\left(A^{(I)1}_{(\alpha_1\cdots\alpha_I)(\dot\alpha_1\cdots\dot\alpha_{I+1})}\right)
     \cdot\left(A^{(I)1}_{(\beta_1\cdots\beta_I)(\dot\beta_1\cdots\dot\beta_{I+1})}
     \eps^{\dot\beta_{I+1}\dot\alpha_{I+1}}\right)
     \cdot\cr
     &
     \quad\cdots\left(A^{(I)1}_{(\omega_1\cdots\omega_I)(\dot\omega_1\cdots\dot\omega_{I+1})}
     \eps^{\dot\omega_{I+1}\dot\alpha_{1}}\eps^{\dot\omega_{I}\dot\beta_{1}}
     \eps^{\dot\omega_{I-1}\dot\gamma_{1}}
     \cdots\right)
\end{align}
with all undoted indices symmetrized.

The above operator can be generalized by stopping the multiplication
before all $\SU(2)_R$ indices are exhausted\footnote{The use of the
$(\bar J_-)$ notation is only schematic.}:
\begin{align}
    \Adp^{(I,\bar I)}
    =\prod_{m=0}^{\bar I}(\bar J_-)^m A^{(I)1}
    &\equiv~\left(A^{(I)1}_{(\alpha_1\cdots\alpha_I)(\dot\alpha_1\cdots\dot\alpha_{I+1})}\right)
     \cdot\left(A^{(I)1}_{(\beta_1\cdots\beta_I)(\dot\beta_1\cdots\dot\beta_{I+1})}
     \eps^{\dot\beta_{I+1}\dot\alpha_{I+1}}\right)
     \cdot\cr
     &\cdots\left(A^{(I)1}_{(\chi_1\cdots\chi_I)(\dot\chi_1\cdots\dot\chi_{I+1})}
     \eps^{\dot\chi_{I+1}\dot\alpha_{I+2-\bar I}}\eps^{\dot\chi_{I}\dot\beta_{I+2-\bar I}}
     \eps^{\dot\chi_{I-1}\dot\gamma_{I+2-\bar I}}
    \cdots\right)
\end{align}
with all uncontracted indices of the same type (doted and undoted)
symmetrized. The operator has exactly $(1+\bar I)$ fermions,
independently of the level I.

We define the generalized open shells as :
\begin{equation}\label{gen-hybrid}
    \prod_{I=0}^{K'-1}\Jdet\left[\Adp^{(I,I+1)}\right]\prod_{I=K'}^K\Jdet\left[\Adp^{(I,K')}\right]
    \left(C^{(K+1)}\Bp^{(\vec J)}\right)_{GI}
\end{equation}
This operator will be BPS if the contractions of the $\SU(2)_R$
indices of the $B$'s are limited in a similar fashion to the fermions
(i.e, $\bar J-\bar J^3$ of each $B$ smaller than $K'/2$). Taking
$K'=\beta K$, the charges of the operator (for simplicity ignoring the
contribution from the bosons) are:
\begin{align}
    &\frac{2J}{N^2}\approx\sum_{I=0}^{\beta K-1}I^2+\sum_{I=\beta K}^{K}I(\beta K+1)
    \approx\frac\beta2\left(1-\frac{\beta^2}{3}\right)K^3\\
    &\frac{2\bar J}{N^2}\approx\sum_{I=\beta K}^{K}(I+1-\beta K)(\beta K+1)
    \approx \frac{\beta}{2}(1-\beta)^2K^3\\
    &\frac{\Qp}{N^2}\approx\sum_{I=0}^{\beta K-1}I+\sum_{I=\beta K}^{K}(\beta K+1)
    \approx \beta\left(1-\frac\beta2\right)K^2
\end{align}
Solving for K, we find,
\begin{equation}
    J=\frac1{\gamma(\beta)}\bar J=
    \alpha(\beta)\frac{N^2\sqrt2}{\sqrt{1-\gamma(\beta)^2}}\left(\frac{\Qp}{N^2}\right)^{3/2}\,,
\end{equation}
with,
\begin{align}
    &\gamma(\beta)=\frac{(1-\beta)^2}{1-\frac{\beta^2}{3}}\,,\cr
    &\alpha(\beta)=\frac{\sqrt{\left(1-\frac23\beta\right)\left(1-\beta+\frac{\beta^2}{3}\right)}}
    {(2-\beta)^{3/2}}\,.
\end{align}
In the allowed range $\beta\in[0,1]$, the ratio $\alpha(\beta)$ is
bounded by $\frac1{2\sqrt2}$, which again is smaller than the
supergravity result \eqref{JtoQ_ratio}.

The shell construction has an interesting scaling property if
we take a small $K'=M$ not scaling with K, i.e all shells are almost
empty. The charges of the operator (for simplicity ignoring
contribution form the B's) are:
\begin{subequations}
\begin{align}
    &\frac{2J}{N^2}\approx\sum_{I=0}^{M-1}I^2+\sum_{I=M}^K (M+1)I\approx(M+1)\frac{K^2}{2}\\
    &\frac{2\bar J}{N^2}\approx\sum_{I=M}^K (I+1-M)(M+1)\approx(M+1)\frac{K^2}{2}-K(M^2-1)\\
    &\frac{\Qp}{N^2}\approx\sum_{I=0}^{M-1}I+\sum_{I=M}^K(M+1)\approx (M+1)K
\end{align}
\end{subequations}
And the ratios are:
 \begin{equation}
    J=\frac{2}{5-b}\,\frac{N^2(1-b)}{4}\left(\frac{\Qp}{N^2}\right)^2
    \csp2
    J-\bar J=\frac{1+3b}{4(1-b)}\Qp
\end{equation}
with $b=\frac{2M-3}{2M+1}$~ defined so that the  $J-\bar J$ equation
matches the supergravity result \eqref{JminBarJ-sugra}. This result
has the asymptotic behavior of the black holes with $J=\bar J$,
missing the supergravity ratio \eqref{JbarJtoQ ratio} by a factor
$~\frac{2}{(5-b)}\leq\frac12$. The open fermi surface described in
section-\ref{z-tit-fermi}, is the $M=0$ case.

\section{Summary and Outlook}\label{z-tit-conc}

In this paper, we used the $\N=4$ field theory at weak non-zero
coupling to reproduce the relations $J(Q)$ in \bps{16} black holes
in $AdS_5\times S^5$ in the regimes \eqref{JtoQ_ratio} and
\eqref{JbarJtoQ ratio}. The main ingredient in our construction is
the filling of fermi surfaces which is used to cancel the
supersymmetry variation of the operator $(D_{\alpha\dot\alpha})^n$.
We expect the fermi sea to play an important role in the microscopic
description of any \bps{16} \ads5 black holes in the limit of large
$J$ and ${\bar J}$ (since the CFT operators will  contain many
covariant derivatives). It would be interesting to study the
3-charge generalization of our discussion, and in particular, to
understand how the complicated angular momentum and R-charges
relations in supergravity appear from the field theory dual for this
cases.

We have used only a subset of the allowed fields and shell
configurations. It is therefore not very surprising that we did not
find the exact $J(Q)$ or $S(Q)$ of the operators as computed in
\cite{Gutowski:2004ez}. We expect that the latter also uses the
fermionic shell structure that we discussed here. If our operators
are indeed primary, as we conjecture, our results suggest the
existence of new \bps{16} black objects in $AdS_5\times S^5$.

There are two lines of generalizations that one can consider. In this work,
we have mostly focused on two specific filling of shells -
one in which the full ${\bar J}$ multiplet is filled, and one in
which only states with near to maximal ${\bar J}_3$ are filled. We
briefly mentioned other possibilities. Clearly there is a rich
variety of allowed fillings and the classification of all possible
$J(Q)$ relations will be carried elsewhere \cite{MJD:2007}.

For example, consider a fermi sea constructed from two regions as
depicted in figure \ref{fig:conc}. Region A in which full ${\bar J}$
multiplets are filled up to some ${\bar J}_0=K'/2$ (i.e., up to
$\sim K'$ derivatives). In region B, from angular momentum
$\frac{K'+1}{2}$ up to some ${\bar J}_1=K/2$ where we fill states
with ${\bar m}={\bar J}$ and ${\bar m}=-{\bar J}$. This state can be
projected to a ${\bar J}=0$ state. This configuration interpolates
between the relation $J/N^2\propto(\Qp/N^2)^{3/2}$ for ${\bar
J}_1-{\bar J}_0\ll{\bar J}_0$ (no region B) and $J/N^2\propto
(\Qp/N^2)^2$ for ${\bar J}_0\ll{\bar J}_1$ (no region A). Of course,
since we have not exhibited a full supersymmetric completion of this
specific mixture of shell filling, we do not know for sure that such
an operator exists, but we find it very plausible.
\FIGURE{\centerline{\vbox{\hspace{3cm}\includegraphics[width=7cm]{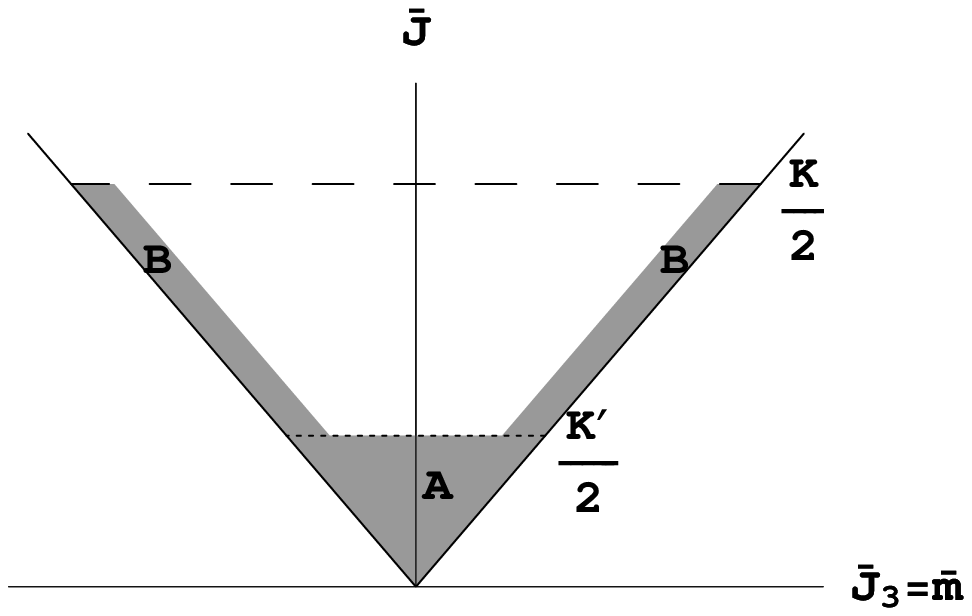}
\caption{Fermi sea of 2 two regions model.\label{fig:conc}}}}}

The interpretation of this state in $AdS_5$ is also unclear. The two
regions $B$, if continued all the way down to ${\bar J}=0$
correspond to two black holes with ${\bar J}=J$ but with opposite
${\bar J}_3$. Region A, if taken by itself, might correspond to a
single black hole with ${\bar J}=0$. What does the full
configuration correspond to ? Does it correspond to highly
deformed black holes, in which the angular momenta ${\bar J}_3$ is
distributed non-uniformly in space ? We believe this to be the case,
although more work is needed to verify this picture \cite{MJD:2007},
but it is clear that we have not exhausted the full range of
possible $J(Q)$ and $S(Q)$ scalings, nor space-time morphologies of
the black holes.

The second possible generalization involves adding more types of fields. A set of
attractive candidates are the field strength operators
$F_{\alpha\beta}$ and theirs derivatives ($G^{(I+2)}$'s). These
operators carry no R-charges. Thus they are excellent candidates to
improve the angular momentum to R-charge ratio reported in this paper.
A BPS combination probably involves the addition of chiral fermions $\l_{\alpha i}$
needed to cancel the $F_{\alpha\beta}$'s supersymmetry
transformations. More precisely, acting with $Q^1$ on $D.....DF$
generates $A$'s from variations of the covariant derivatives, for
which we need shells as we discussed so far, and
$D....D[\lambda_{i\alpha},M^{1i}]$. Including the latter in the
operator from the start means that the SUSY variation of $D....DF$
will be zero within this operator (the $Q^1$ variation of
$[\lambda_{i\alpha},M^{1i}]$ is zero). It is interesting to point out the existence of
supersymmetric $AdS_5\times S^5$ configurations having angular momentum but no R-charge \cite{mirjamjoan}.
The existence of this type of operators, which is left to future work, could provide evidence for the
existence of these spacetimes in string theory, since the existence of a naked singularity of the latter render
their interpretation unclear.

\section{Acknowledgments}\label{z-tit-ack}

We would like to thank O. Aharony, Y. Antebi, D.Kutasov, F. Larsen
and S.Minwalla for useful discussions and comments. JS would like to
thank the Weizmann Institute of Science for hospitality during
different periods in the completion of this work. The work of MB is
supported by the Israel Science Foundation, by the Braun-Roger-Siegl
foundation, by EU-HPRN-CT-2000-00122, by GIF, by Minerva, by the
Einstein Center and by the Blumenstein foundation. JS is supported
in part by the DOE under grant DE-FG02-95ER40893, by the NSF under
grant PHY-0331728 and by an NSF Focused Research Grant DMS0139799.

\appendix

\section{A Note on $\SO(6)$ Representations}\label{z-tit-apd-a}

In the supergravity literature, the standard choice of simple roots
and fundamental weight of the $\SO(6)$ algebra is
\begin{align}
    & \tilde\alpha^1=\frac1{\sqrt2}\left(0,1,1\right) &
    & \tilde\mu^1=\frac1{\sqrt8}\left(1,1,1\right)~=~{\Yboxdim6pt\yng(1)} &
    \cr
    & \tilde\alpha^2=\frac1{\sqrt2}\left(1,-1,0\right) &
    & \tilde\mu^2=\frac1{\sqrt8}\left(2,0,0\right)~=~{\Yboxdim6pt\yng(1,1)}  &
    \cr
    & \tilde\alpha^3=\frac1{\sqrt2}\left(0,1,-1\right) &
    & \tilde\mu^3=\frac1{\sqrt8}\left(1,1,-1\right)~=~{\Yboxdim6pt\yng(1,1,1)}  &
\end{align}
A representation can be expressed using a Young tablea with $k,p,q$
columns of heights $1,2,3$ respectively. The highest weight of the
representation is
\begin{equation*}
    \tilde\mu=\frac1{\sqrt8}\left(k+2p+q~,~k+q~,~k-q\right)
\end{equation*}

When we discuss the $\N=4$ SYM, we follow the notation of
\cite{Dolan:2002zh} using the Dynkin labels of $\SO(6)\cong\SU(4)$.
The related choice of simple roots and fundamental weights:
\begin{align}
    & \hat\alpha^1=\left(2,-1,0\right) &
    & \hat\mu^1=\left(1,0,0\right)~=~{\Yboxdim6pt\yng(1)} & \cr
    & \hat\alpha^2=\left(-1,2,-1\right) &
    & \hat\mu^2=\left(0,1,0\right)~=~{\Yboxdim6pt\yng(1,1)}  & \cr
    & \hat\alpha^3=\left(0,-1,2\right) &
    & \hat\mu^3=\left(0,0,1\right)~=~{\Yboxdim6pt\yng(1,1,1)}  &
\end{align}
In the Dynkin labels the highest weights of a representations are
identical to the number of columns of each height ($k,p,q$).

Comparing the above, the translation between the supergravity
notations ($Q_i$) and the $\N=4$ notations ($k,p,q$) is:
\begin{equation}
    Q_1=\frac{k+2p+q}{2 l}\qquad
    Q_2=\frac{k+q}{2 l}\qquad
    Q_3=\frac{k-q}{2 l}
\end{equation}
The overall factor is set by matching the $\N=4$ and supergravity
BPS formula's.
%

%
%
%

\section{Charges in The Fermionic Bands
Model}\label{z-tit-apd-fermion}

We start with the single fermionic band model described by the
operator:
\begin{equation}
    \prod_{I\in R}\prod_{m=0}^{I+1}\Jdet\left[(\bar
    J_-)^mA^{(I)}\right]\left(C^{(K+1)}\Bp^{(\vec J)}\right)_{GI}
\end{equation}
with $J_i\leq K+1$ and where the set R is defined by :
\begin{equation}
    R=\left\{r\in\mathbb{Z}\bigm|0\leq r\leq K ~\cup~ Ks\leq r\leq 2K
    \right\}\quad
    s\in[1,2]
\end{equation}
The quantum numbers of this operator in the large angular momentum and R-charge limit are:
\begin{align}
    &\frac{J}{N^2}\leq\frac{K^3}{6}+\frac{(2K)^3-(Ks)^3}{6}+\frac{3 KL}{2N^2}\,,\cr
    &\frac{\Qp}{N^2}=\frac{K^2}{2}+\frac{(2K)^2-(Ks)^2}{2}+\frac{2L}{N^2}\,.
\end{align}
Using the second equation to eliminate $L$, we find the relation:
\begin{equation}
    \frac{J}{N^2}\leq\alpha(y,s){\sqrt 2}\left(\frac{\Qp}{N^2}\right)^{3/2}
\end{equation}
with, $y\equiv K\Bigm/\sqrt{\frac{\Qp}{N^2}}$~ and
\begin{equation}
    \alpha(y,s)\equiv
    \frac3{4\sqrt2}
    y\left[1-\frac12y^2+\frac12y^2\left(s^2-\frac{4}{9}s^3\right)\right]\,.
\end{equation}
Maximizing $\alpha$ over $\{y,\,s\}$, keeping in mind that $L\geq0$,
the solution is found on the boundary of the allowed range with
$L=0$ (i.e only fermions):
\begin{equation}
    \alpha_{max}=\alpha(y,s)\bigm|_{y=\frac{5}{2\sqrt{11}},s=\frac95}=
    \frac{3}{2\sqrt{11}}\approx0.45.
\end{equation}
The above example demonstrates a property common to operators with
fermionic bands : the best ratio (maximal $\alpha$) is found when all
the angular momentum comes from the fermions ($L=0$).

Having this experience, we look for the best configuration of
fermions in $n$ bands. We start by occupying all fermions up to
level $(n+1)K$, then removing fermions in the level's range $\left(m
K, (\frac32-s)m K\right)$, with $m$ an integer smaller than $(n+1)$
and $s$ a real number in the range
$\left[\frac{n-2}{2n},\frac12\right]$. The maximal and minimal
values come from the condition that there are some fermions in the
upper band:
\begin{equation}
    nK<(\frac32-s)nK<(n+1)K\,.
\end{equation}
The charges in the large angular momentum and R-charge limit are :
\begin{align}
    \frac{J}{N^2}=&\frac{K^3}{6}\sum_{m=0}^{n}\left[(m+1)^3-\left(\frac32-s\right)^3m^3\right]+\frac{3K L}{2N^2}\,,\\
    \frac{\Qp}{N^2}=&\frac{K^2}{2}\sum_{m=0}^{n}\left[(m+1)^2-\left(\frac32-s\right)^2m^2\right]+\frac{2L}{N^2}\,.
\end{align}
Repeating the same procedure as above, we find:
\begin{equation}
    \frac{J}{N^2}=\alpha(y,s;n){\sqrt2}\left(\frac{\Qp}{N^2}\right)^{3/2}\,.
\end{equation}
Once again, the maximal value for $\alpha$ is found when $L=0$:
\begin{equation}
    \alpha_{max}(n)=\frac{\sqrt3}{2}\sqrt{\frac{(n+1)^2-1}{4(n+1)^2-5}}\,.
\end{equation}
$\alpha_{max}(n)$ is a monotonically decreasing function, and we
conclude that the upper bound is for $n=1$, lower than the
supergravity constraint by a factor of $\approx0.45$.

\end{document}